\documentclass[a4paper,11pt]{article}
\usepackage{amsmath}
\usepackage{amsthm}
\usepackage{amssymb}
\usepackage{graphicx}
\usepackage{xcolor}
\usepackage{dsfont}
\usepackage{array}
\usepackage{bm}

\definecolor{darkgreen}{HTML}{32CD32}

\def\defd#1{{\bf #1}}

\newcount\prevparpenalty
\makeatletter
\def\maxitemizepenalty{\prevparpenalty=\@beginparpenalty\@beginparpenalty=10000}
\def\restoreitemizepenalty{\@beginparpenalty=\prevparpenalty}
\makeatother

\def\delim#1{%
\noindent%
\leavevmode%
\raise.3em\hbox to\hsize{%
\lower.3em\hbox{\vrule height.3em}%
\hrulefill\ %
\lower.3em\hbox{{\bf #1}}\ %
\hrulefill%
\lower.3em\hbox{\vrule height.3em}}%
\par\penalty10000%
}
\def\enddelim{%
\penalty10000\par\penalty10000%
\noindent%
\leavevmode%
\raise.3em\hbox to\hsize{%
\vrule height.3em%
\hrulefill%
\vrule height.3em}%
\par%
}

\newtheorem{theorem}{Theorem}
\newtheorem{proposition}[theorem]{Proposition}
\newtheorem{lemma}[theorem]{Lemma}
\newtheorem{definition}[theorem]{Definition}
\newtheorem{corollary}[theorem]{Corollary}

\numberwithin{equation}{section}
\numberwithin{theorem}{section}
\numberwithin{figure}{section}

\begin{document}

\title{A Pfaffian formula for monomer-dimer\\ partition functions}

\author{\vspace{5pt} Alessandro Giuliani$^{1}$, Ian Jauslin$^{2}$ and Elliott H.~Lieb$^{3}$\\
  \vspace{-4pt}\small{$^{1}$Dipartimento di Matematica e Fisica,  Universit\`a di Roma
Tre} \\ \small{
L.go S. L. Murialdo 1, 00146 Roma, Italy}\\
  \vspace{-4pt}\small{$^{2}$Dipartimento di Fisica, Sapienza Universit\`a di Roma} \\ \small{
P.le A. Moro 5, 00185 Roma, Italy}\\
  \vspace{-4pt}\small{$^{3}$Departments of Mathematics and Physics,
    Jadwin Hall, Princeton University}\\
\small{Princeton 08544 NJ, USA}}

\date{January 5, 2016}

\maketitle

\begin{abstract}
We consider the monomer-dimer partition function on arbitrary finite planar graphs and arbitrary monomer and dimer weights, with the restriction that 
the only non-zero monomer weights are those on the boundary. We prove a Pfaffian formula for the corresponding partition function.
As a consequence of this result, multipoint boundary monomer correlation functions at close packing are shown to satisfy fermionic statistics. 
Our proof is based on the celebrated Kasteleyn theorem, combined with a theorem on Pfaffians proved by one of the authors, and a careful 
labeling and directing procedure of the vertices and edges of the graph.
\end{abstract}

{\small
\tableofcontents
}

\section{Introduction}

The monomer-dimer problem is one of the important classical structures in statistical mechanics and computer science. It starts with a {\it graph} $g$, which is a collection of points, called {\it vertices}, and lines, called {\it edges}, between specified pairs of points. A dimer is an object that occupies a single edge and its endpoints, and a monomer is an object that occupies a single vertex. A {monomer-dimer covering} of $g$ (hereafter refered to as an MD covering) is a collection of monomers and dimers (that is to say vertices and edges) such that every vertex is covered by exactly one of these objects, that is by either a monomer or a dimer. Note that, in any MD covering, the number of vertices in $g$ is equal to the number of monomers plus twice the number of dimers. The present work is devoted to {\it planar} graphs (which are those that can be drawn in $\mathbb R^2$ without edge crossings).

The classical problem associated with MD overings is their enumeration at fixed number of monomers. The object of this work is to find formulas for the generating function of this enumeration problem:
\begin{equation}
\Xi(z):=\sum_{\mathrm{MD\ coverings}}z^{\mathrm{number\ of\ monomers}}.
\label{eqXizdef}\end{equation}
Clearly, $\Xi(z)$ is a polynomial in $z$, and $z$ is called the {\it monomer fugacity}. This polynomial has all its zeros on the imaginary axis~\cite{HL70,HL72}. In addition, the summation in~(\ref{eqXizdef}) can be generalized by assigning weights to edges and/or vertices.

In the pure dimer case, where $z=0$, $\Xi$ has been shown by Temperley and Fisher (for the square lattice) \cite{TF61}  and by Kasteleyn (for general planar graphs) \cite{Ka63} to be expressible as a Pfaffian (which is convenient since Pfaffians can be computed as square roots of determinants). However, when monomers are allowed to appear, such a Pfaffian formula is thought to be impossible (at least a Pfaffian formula for the {\it full} MD problem on {\it any} planar graphs): indeed it has been shown \cite{Je87} that the enumeration of MD coverings of generic planar graphs is ``computationally intractable'', whereas Pfaffians can be computed in polynomial time. More precisely, \cite{Je87} proves that the enumeration of MD coverings of generic planar graphs is
``$\# P$ complete'', which implies that it is believed not to be computable in polynomial time.

However, by introducing restrictions on the location of monomers, such a result can be proven in some cases. Namely, in \cite{TW03,Wu06}, the authors derive a Pfaffian formula, based on the ``Temperley bijection'' \cite{Te74}, for the partition function of a system with a {\it single} monomer located on the boundary of a finite square lattice, and in \cite{WTI11}, on a cylinder of odd width (which is a nonbipartite lattice). In \cite{PR08}, the MD problem is studied on the square lattice on the half-plane with the restriction that the monomers are {\it fixed} on points of the boundary. They derive a Pfaffian formula for this case, and use it to compute the scaling limit of the multipoint boundary monomer correlations. Finally, in \cite{AF14}, it is shown that if the monomers are {\it fixed} at any position in a square lattice, then the partition function can also be written as a product of two Pfaffians.
\bigskip

In the present work, we prove a Pfaffian formula for the {\it boundary} MD partition function on an {\it arbitrary} planar graph (in which the monomers are restricted to the boundary of the graph, but are not necessarily fixed at prescribed locations) with arbitrary dimer and monomer weights.

It was later brought to our attention by an anonymous referee that it is known that the boundary MD partition function can be given by {\it a} Pfaffian formula, which one can obtain by considering a bijection between the boundary MD coverings of a graph and the pure dimer coverings of a larger graph, whose partition function is known, by Kasteleyn's theorem~\cite{Ka63}, to be expressible as a Pfaffian. This construction is detailed in appendix~\ref{appbijectionmethod}. To our knowledge, this result has, so far, not been published, even though it appears to be closely related to the discussion in \cite[section 4]{Ku94}.

Oblivious to the existence of this bijection method, the approach we have adopted in this paper is a different one. Instead of mapping the boundary MD coverings to pure dimer coverings of a larger graph, we compute, for each fixed monomer positions, the pure dimer partition function on the subgraph obtained by removing the vertices covered by monomers, which, by Kasteleyn's theorem, is expressible as a Pfaffian, and combine all the Pfaffians thus obtained into a single Pfaffian using a theorem proved by one of the authors in 1968~\cite{Li68}. The formula we obtained in this way is slightly different from that obtained by the bijection method, and we have found that, by using this formula, the monomer correlations functions at close packing can easily be shown to satisfy a fermionic Wick rule.

In short, the aim of this paper is, first, to leave a written trace of the fact that the boundary MD partition function can be expressed as a Pfaffian, and, second, to present a Pfaffian formula, which is not a trivial rewriting of the formula one obtains by the bijection method, and may be easier to use than the latter formula in some cases, for instance in the proof of the Wick rule.

Finally, it should be mentioned that, by drawing inspiration from the bijection method detailed by our anonymous referee, we found a significant simplification of our proof, and we are, therefore, very grateful.
\bigskip

{\bf Remarks}:
\maxitemizepenalty
\begin{itemize}
\item The asymptotic behavior of monomer pair correlations on the square lattice have been computed explicitly \cite{FS63, Ha66, FH69} for monomers on a row, column or diagonal (note that, as mentioned in~\cite{AP84}, \cite{Ha66} contains small mistakes). In addition, the general bulk monomer pair correlations have been shown~\cite{AP84} to be expressible in terms of two critical Ising correlation functions.
\item An alternative approach for the boundary MD problem on an $N\times M$ rectangle (by which we mean $N$ vertices times $M$ vertices) with monomers allowed only on the upper and lower sides 
would be to use the transfer matrix technique \cite{Li67}. In that case, the boundary MD partition function is written as $x_M\cdot V^{M-1}x_1$ where $V$ is the
($N\times N$) transfer matrix and $x_1$ and $x_M$ are vectors determined by the boundary condition at the boundaries $y=1$ and $y=M$ respectively. 
Since the monomers are only allowed on these two boundaries, the matrix $V$ is the transfer matrix for {\it pure} dimer coverings, which can be 
diagonalized as in \cite{Li67}. The partition function can then be computed by setting the vectors $x_1$ and $x_M$ appropriately. 
\item The boundary monomer correlations at close-packing are critical, in that if the graph is ``regular enough'' (e.g., 
if it is a finite portion of a lattice) they decay polynomially at large distances, 
like $1/(distance)$, asymptotically as the size of the graph tends to infinity. 
See \cite{PR08} for a proof of this fact on the square lattice on the half-plane. 
A similar analysis has been worked out in the 2D nearest neighbor Ising model for the boundary free energy, in the presence of a boundary magnetic field, and for the boundary spin-spin correlations, see \cite[Section 8]{MW67} and \cite[Chapters VI and VII]{MW73}.
If the graph is a discrete, regular approximation of a finite domain of $\mathbb R^2$, 
the scaling limit of the boundary monomer correlations at close-packing
is expected to exist and to be conformally invariant under conformal mappings of the domain, in analogy with other observables 
of the critical 2D Ising model and of the close-packed dimer model \cite{Ke00,Ke01, Sm01,Sm10,CHI15,Du11,Du15}.
In particular, they are expected to coincide with those of complex chiral free fermions \cite{PR08}.
It is unclear whether this scaling limit is stable under perturbations violating planarity (e.g., under the addition of small dimer weights along extra 
edges crossings). Our Pfaffian formula offers a starting point for a perturbative multiscale analysis of the problem, in the spirit of \cite{PS,Sp00,GGM12, GMT15b, GMT15}.
\item An alternative approach for the boundary MD problem on generic
planar graphs is via the random current representation developed by
Aizenman \cite{Ai82}. It has been recently observed
\cite{AD} that this representation, adapted to planar lattices, implies, for
purely geometrical
reasons, the validity of the fermionic Wick rule for boundary spin
correlations in the nearest neighbor Ising model (which has already been proved by J.~Groeneveld, R.J.~Boel and P.W.~Kasteleyn~\cite{GBK78}), and might imply the
same for boundary monomer correlations in the dimer model.
Their method also suggests a stochastic geometric perspective on the
emergence of planarity at the critical points of non-planar 2D models, in the sense of the previous item. Note that our Pfaffian formula (see theorem~\ref{theomain}) goes beyond the Wick rule, see the remark at the end of section~\ref{subsecpfaffian}.
\item It may be worth noting that the MD partition function can be computed exactly in some cases, e.g. on the complete graph~\cite{ACM14}.
\end{itemize}
\restoreitemizepenalty

We will now state our main result more precisely, for which we need some notation. Let $\mathcal G$ denote the set of finite planar graphs with edge-weights and vertex-weights, embedded in $\mathbb R^2$, that have an even number of vertices, and contain no {\it double edges} or {\it self-contractions} (that is the endpoints of an edge are distinct, and no two edges share the same endpoints). Note that these graphs are not necessarily connected. The 
evenness condition is not restrictive, in that a graph with an odd number of vertices can always be reduced to an even one, by adding an isolated (disconnected) vertex.

Given $g\in\mathcal G$, its \defd{boundary graph} $\partial g$ is defined as the sub-graph of $g$ containing the edges and vertices that can be connected to infinity without crossing any edge of $g$ (here we say that an edge can be connected to infinity without crossing any other edge, if a point at the center of the edge can be). The set of vertices of $g$ is denoted by $\mathcal V(g)$ and its set of edges by $\mathcal E(g)$. The edge linking two vertices $v_1$ and $v_2$ will be denoted by $\{v_1,v_2\}\equiv\{v_2,v_1\}$. The weight of a vertex $v\in\mathcal V(g)$ (i.e. the fugacity of a monomer located at $v$) is denoted by $\ell_v$ and the weight of an edge $\{v_1,v_2\}\in\mathcal E(g)$ (i.e. the fugacity of a dimer located at $\{v_1,v_2\}$) is denoted by $d_{\{v_1,v_2\}}\equiv d_{v_1,v_2}$. The number of vertices in $g$ in denoted by $|g|$. 
In the following we will often consider {\it directed} graphs, which are obtained by assigning a direction to every edge: if the edge $\{v_1,v_2\}$ is directed from $v_1$ to $v_2$, we write $v_1\succ v_2$.

The set of MD coverings of $g$ is denoted by $\Omega(g)$ and the set of pure dimer coverings by $\Omega_0(g)$. Given an MD covering $\sigma\in\Omega(g)$, we denote the set of vertices covered by monomers by $\mathfrak M(\sigma)\subset\mathcal V(g)$ and the set of edges covered by dimers by $\mathfrak D(\sigma)\subset\mathcal E(g)$. The \defd{boundary MD partition function} of a graph $g$ is defined as the partition function of MD coverings of $g$ in which the monomers are restricted to vertices of $\partial g$:
\begin{equation}
\boxed{\ \Xi_\partial(\bm\ell,\mathbf d):=\sum_{\displaystyle\mathop{\scriptstyle\sigma\in\Omega(g)}_{\mathfrak M(\sigma)\subset\mathcal V(\partial g)}}\prod_{v\in\mathfrak M(\sigma)}\ell_v\prod_{e\in\mathfrak D(\sigma)}d_e.\ }
\label{eqpartfnboundary}\end{equation}
Note that restricting the monomers of $\sigma$ to be on boundary vertices can be enforced by setting all other $\ell_v$'s to 0.

The Pfaffian of a $2n\times2n$-dimensional antisymmetric matrix $(A_{i,j})$
is defined as
\begin{equation}
 \mathrm{pf}(A):=\frac1{2^nn!}\sum_{\pi\in\mathcal
S_{2n}}(-1)^\pi\prod_{i=1}^nA_{\pi(2i-1),\pi(2i)}
\label{eqpfaffdef}\end{equation}
where $\mathcal S_{2n}$ denotes the set of permutations of $\{1,\cdots,2n\}$ and $(-1)^\pi$ is the signature of $\pi\in\mathcal S_{2n}$.

\begin{theorem}[Main result]\label{theomain}
For every $g\in\mathcal G$, the edges of $g$ can be directed and its vertices labeled $(v_1,\cdots,v_{|g|})$ in such a way that, by defining
\begin{equation}
a_{i,j}(\mathbf d):=\left\{
\begin{array}{l@{\ }l}
 +d_{v_i,v_j}&\mathrm{if\ }\{v_i,v_j\}\in\mathcal E(g)\ \mathrm{and\ }v_i\succ v_j\\
 -d_{v_i,v_j}&\mathrm{if\ }\{v_i,v_j\}\in\mathcal E(g)\ \mathrm{and\ }v_i\prec v_j\\
 0&\mathrm{otherwise}
\end{array}
\right.
\label{eqAdef}\end{equation}
and, for $i<j$,
\begin{equation}
A_{i,j}(\bm\ell,\mathbf d):=a_{i,j}(\mathbf d)-(-1)^{i+j}\ell_i\ell_j
\label{eqpolysta}\end{equation}
and $A_{j,i}(\bm\ell,\mathbf d):=-A_{i,j}(\bm\ell,\mathbf d)$, the boundary MD partition function on $g$ is given by
\begin{equation}
\boxed{\ \Xi_\partial(\bm\ell,\mathbf d)=\mathrm{pf}(A(\bm\ell,\mathbf d)).\ }
\label{eqtheo}\end{equation}
\end{theorem}
\bigskip

\noindent {\bf Remarks}:
\maxitemizepenalty
\begin{itemize}
\item By setting some weights $\ell_v$ to 0, the location of the monomers can be further restricted. In particular, the partition function (and correlation functions) with {\it fixed} monomers can be expressed using the Pfaffian formula~(\ref{eqtheo}), as has been done for the special cases studied in \cite{TW03,PR08}.
\item Similarly, by setting edge weights to 0, dimers can be excluded from any part of the graph.
\item It may be worth noting that the weights $d_e$ and $\ell_v$ may be complex.
\item Our result provides a polynomial-time algorithm for computing boundary MD partition functions on generic planar graphs. 
See appendix \ref{app.A} for examples. 
\item Based on theorem \ref{theomain}, we have derived an algorithm (see appendix~\ref{appalg}) that allows us to compute the {\it full} MD partition function (as opposed to the boundary MD partition function) on an arbitrary graph (which is not necessarily planar), which is more efficient than the naive enumeration algorithm. 
For instance, if $g$ is an $L\times M$ rectangle on the square lattice, our algorithm requires 
$O((LM)^3 2^{(LM)/2})$ operations, while the naive algorithm requires $O((LM)^32^{LM})$. In the rectangular case, a 
transfer matrix approach would be even faster, completing the computation in $O((LM)^32^L)$ operations~\cite{Ko06c}, but our algorithm 
does not require the graph to be treatable via a transfer matrix approach. 
\item In addition, we have developed an alternative algorithm (see appendix~\ref{appinout}) to express the {\it full} MD partition function on Hamiltonian planar graphs as a derivative of the product of just two Pfaffians. From a computational point of view, this approach is even slower than the previous one, but it is nonetheless conceptually interesting. Note that this algorithm can be adapted to non-planar graphs as well.
\item Finally, we have also computed upper and lower bounds for the {\it full} partition function, see theorems~\ref{theolowerbound} and~\ref{theoupperbound}.
\item As a side remark, note that Monte Carlo methods methods can be even faster, i.e., polynomial in the size of the system~\cite{KRS96}, but they provide correct results only with high probability rather than with certainty.
\item A result similar to theorem~\ref{theomain} has recently been established~\cite{Ay15} for another model, called the {\it monopole-dimer} model, for which the partition function can be written as a determinant.
\end{itemize}
\restoreitemizepenalty
\bigskip

If we derive $\Xi_\partial(\bm\ell,\mathbf d)$ with respect to $\bm \ell$ and then set $\bm\ell$ to zero, we obtain the multipoint 
monomer correlations at close packing:
\begin{equation}
M_n(i_1,\cdots,i_{2n}):=\left.\frac1{\Xi_\partial(\mathbf0,\mathbf d)}\frac{\partial^{2n}\Xi_\partial(\bm\ell,\mathbf d)}{\partial\ell_{i_1}\cdots\partial\ell_{i_{2n}}}\right|_{\ell_1=\cdots=\ell_{|g|}=0}.
\label{eqcorrdef}\end{equation}
An important corollary of theorem \ref{theomain} is that $M_n(i_1,\cdots,i_{2n})$ satisfies the fermionic Wick rule.
\begin{corollary}[Fermionic Wick rule]\label{corrmain}
In the same setting as theorem \ref{theomain},
\begin{equation}
M_n(i_1,\cdots,i_{2n})=\mathrm{pf}(\mathfrak M_{i_1,\ldots,i_{2n}}),
\label{eqwick}\end{equation}
where $\mathfrak M_{i_1,\ldots,i_{2n}}$ is the $2n\times 2n$ antisymmetric matrix whose $(j,j')$-th entry with $j<j'$ is $M_1(i_j,i_{j'})$.
\end{corollary}

\noindent{\bf Remark}: Away from close packing (i.e. omitting $\ell_1=\cdots=\ell_{|g|}=0$ in~(\ref{eqcorrdef})), {\it the Wick rule does not hold}. This can be checked immediately by considering a graph consisting of a square with an extra edge on the diagonal.

\bigskip

As stated in theorem~\ref{theomain}, the edges and vertices of $g$ must be directed and labeled in a special way. In particular, the direction of the edges must satisfied a so called {\it Kasteleyn} condition, and the labeling must satisfy a {\it positivity} condition. The positivity condition ensures that the terms that appear in the Pfaffian add up constructively and reproduce the MD partition function. The Kasteleyn condition is used to prove the positivity of a graph: if such a condition holds, then it suffices to look at a single dimer covering of $g$ to prove its positivity.

The main ingredient of the proof of our result is to show that, having directed and labeled the graph in an appropriate way, every sub-graph constructed from $g$ by removing the vertices which support monomers satisfies both the Kasteleyn and positivity conditions. Proving that the sub-graph satisfies the Kasteleyn condition is easy (provided the monomers live on the boundary of the graph), but proving its positivity is more of a challenge. The basic idea is to construct an auxiliary graph in which the boundary MD coverings of $g$ are mapped to dimer coverings by a map that preserves positivity. We can then show that the auxiliary graph is positive, which implies the positivity of the sub-graphs of $g$.
\bigskip

The structure of the paper is the following.
\begin{itemize}
\item In section~\ref{secpreliminaries}, we define and discuss some of the ingredients of the proof of the Pfaffian formula. Namely, we define the Kasteleyn and positivity conditions and state a theorem on Pfaffians, based on a result of~\cite{Li68}, which is at the basis of the Pfaffian formula for the boundary MD partition function. Moreover, we prove corollary \ref{corrmain}. 
\item In section~\ref{secenclosed}, we prove theorem~\ref{theomain} for a class of graphs called {\it enclosed graphs}.
\item In section~\ref{sec7}, we show how to add edges and vertices to a graph in a way that does not change the partition function and reduces it to an enclosed graph.
\item In appendix~\ref{apppropkasteleyn}, we state some useful properties of Kasteleyn graphs, and prove some of the statements of section~\ref{secpreliminaries}.
\item In appendix~\ref{app.A}, we give several examples of the Pfaffian formula.
\item In appendices~\ref{appalg} and~\ref{appinout} we present two algorithms to compute the {\it full} monomer-dimer partition function on arbitrary planar graphs.
\item In appendix~\ref{appbijectionmethod}, we discuss the bijection method.
\end{itemize}
\bigskip

\section{Preliminaries}\label{secpreliminaries}
In this section, we present the key results that will allow us to prove the
Pfaffian formula for the boundary MD partition function.
\bigskip

\subsection{Kasteleyn's theorem}\label{subseckasteleyn}
In this section, we discuss a method introduced by P.~Kasteleyn \cite{Ka63}
to write the partition function of dimers on planar graphs as a Pfaffian.
In order to construct the matrix $A$ whose Pfaffian yields the partition
function of dimers, the graph $g$ must first be oriented and its vertices
labeled in a way that satisfies two conditions: the {\it Kasteleyn} and
{\it positivity} conditions described below. 

\subsubsection{The Kasteleyn condition}
Before discussing the Kasteleyn condition, we first define a
\defd{counterclockwise circuit} $c=(v_1,\cdots,v_{|c|})$ with $|c|\ge 3$ as an ordered sequence of
vertices $v_i\in\mathcal V(g)$ that are such that
\begin{itemize}
\item $v_i\neq v_j$ for all $i\neq j$,
\item for all $1\le i\le |c|$, $\{v_i,v_{i+1}\}\in\mathcal E(g)$, where
$v_{|c|+1}\equiv v_1$,
\item the path $v_1\rightarrow\cdots\rightarrow v_{|c|}\rightarrow v_1$ 
winds in the counterclockwise direction.
\end{itemize}
Note that the notion of counterclockwise circuit does not require the graph to be 
directed. 
The ``counterclockwise'' adjective will be omitted in the following. Moreover, 
we will (obviously) identify circuits obtained from each other by a cyclic permutation.

The \defd{boundary} of a graph $g$ is the set of vertices and edges that
are accessible from infinity. A graph is said to have a
\defd{boundary circuit} if its boundary forms a circuit. Note that all finite
graphs have a boundary, but not always a boundary circuit (e.g. two
vertices connected by an edge). In this paper we will first be concerned
with graphs that have a boundary circuit, and in section~\ref{secboundarycircuit} we will
show how to reduce general graphs to graphs with a boundary circuit. 

Given an edge $\{v_i,v_{i+1}\}$ for $1\le i\le |c|$,
the edge is said to be \defd{forwards} if $v_i\succ v_{i+1}$ and
\defd{backwards} if $v_i\prec v_{i+1}$; and similarly for 
$\{v_{|c|},v_1\}$. A circuit $c$ is said to be \defd{oddly-directed} if it contains an {\it odd} number of {\it backwards} edges and \defd{evenly-directed} if it contains an {\it even} number of {\it backwards} edges. In addition a circuit is said to be \defd{good} if it is {\it oddly-directed} and encloses an {\it even} number of vertices, or it is {\it evenly-directed} and encloses an {\it odd} number of vertices.

\begin{figure}
\hfil\includegraphics{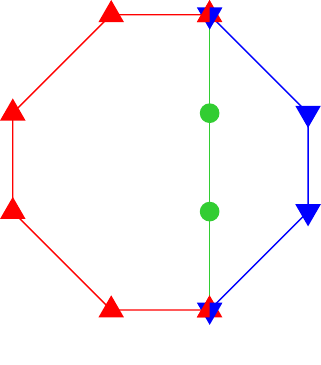}
\caption{A merger. The circuit $c_1$ consists of the vertices rendered as {\color{red}red} (color online) upward-pointing triangles and {\color{darkgreen}green} circles, and the edges connecting them. The circuit $c_2$ consists of the vertices rendered as {\color{blue}blue} downward-pointing triangles and {\color{darkgreen}green} circles, and the edges connecting them. The merger $c_1\Delta c_2$ consists of the vertices rendered as {\color{red}red} upward-pointing triangles and {\color{blue}blue} downward-pointing triangles, and the edges connecting them. The vertices that are rendered as superimposed upward- and downward-pointing triangles, half red and half blue, should be interpreted as {\it both} red upward-pointing triangles {\it and} blue downward-pointing triangles.}
\label{figmerger}
\end{figure}

\begin{figure}
\hfil\includegraphics{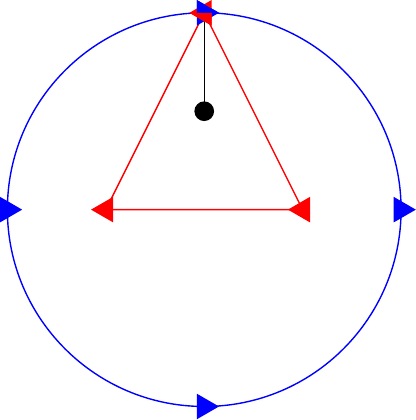}
\caption{In this example, there are two minimal circuits. The first minimal circuit consists of the vertices rendered as {\color{red}red} (color online) left-pointing triangles, and the edges connecting them. The second minimal circuit consists of the vertices rendered as {\color{blue}blue} right-pointing triangles, and the edges connecting them. The vertex that is rendered as superimposed left- and right-pointing triangles, half red and half blue, should be interpreted as {\it both} a red left-pointing triangle {\it and} a blue right-pointing triangle.}
\label{figminimalcircuit}\end{figure}

Furthermore, given $\nu\ge 1$ and two circuits $c_1$ and $c_2$ that have a
string of vertices in common appearing in the reverse order, that is
\begin{equation}
c_1=(v_1,\cdots,v_{\nu+1},v_{\nu+2},\cdots,v_{|c_1|}),\quad
c_2=(v_{\nu+1},\cdots,v_{1},v_{\nu+2}',\cdots,v_{|c_2|}')
\label{eqmergecircuit}\end{equation}
with $v_i\neq v_j$ for all $i\neq j$ and $v_i\neq v'_j$ for all $i,j$.
The edges $\{v_i,v_{i+1}\}$ with $i\le\nu$ are
the edges that $c_1$ and $c_2$ share. We define the \defd{merger} of $c_1$
and $c_2$ as the circuit
\begin{equation}
c_1\Delta
c_2:=(v_{\nu+1},\cdots,v_{|c_1|},v_1,v_{\nu+2}',\cdots,v_{|c_2|}').
\label{eqsymdiff}\end{equation}
See figure~\ref{figmerger} for an example.
A circuit $c$ is said to be \defd{minimal} if it is not a merger, that is if
for any pair of circuits $c_1$ and $c_2$ as in~(\ref{eqmergecircuit}),
$c\neq c_1\Delta c_2$. A circuit $c_1$ is said to be \defd{maximal} if it
cannot be merged with any other circuit, that is if there is no $c_2$ as
in~(\ref{eqmergecircuit}). Note that a minimal circuit may have vertices and edges (and even circuits) in its interior, see figure~\ref{figminimalcircuit} for an example.
Vice versa, a circuit without vertices in its interior is minimal. Minimal circuits with no interior vertices are called {\it mesh cycles} in \cite{Ka63}.

\begin{definition}\label{defkasteleyn}
A directed graph $g\in\mathcal G$ is said to be \defd{Kasteleyn} if every
minimal circuit of $g$ is good, or if there are no minimal circuits in $g$.
\end{definition}
\bigskip

The adjective ``minimal'' can be dropped from definition~\ref{defkasteleyn}, as shown in the following lemma, which we will prove in appendix~\ref{apppropkasteleyn}.
\begin{lemma}\label{lemmakasteleyncomplete}
Every circuit of a Kasteleyn graph is good.
\end{lemma}
\bigskip

\noindent{\bf Remark}: Note that, although our definition is slightly different from that used originally by Kasteleyn in \cite{Ka63}, it can easily be 
recognized to be equivalent. In fact, the assumption of \cite[item (A) on p.290]{Ka63}, in light of \cite[item (D) on p.290]{Ka63}, 
is equivalent, in our language, to the fact that all even circuits are good (here, ``even'' refers to the number of vertices in the circuit, and is unrelated to the notion of ``evenly-directed'' defined above). Moreover, \cite[item (C) on p.290]{Ka63} guarantees that 
both even and odd circuits are good. 

\medskip

An important result of \cite{Ka63} is that every finite planar graph can be directed in such a way that it is Kasteleyn. 
A simple directing procedure alternative to that proposed in \cite{Ka63} can be found in \cite{LL93}. 
We will actually need a slightly generalized version of this directing procedure, 
which applies to graphs that are partially directed.
\begin{proposition}\label{propdirectkasteleyn}
Let $g\in\mathcal G$ be a graph some of whose edges may be directed. If
every circuit that is thus directed is good, then the undirected
edges of $g$ can be directed in such a way that the resulting graph is
Kasteleyn.
\end{proposition}
For a proof and an algorithmic construction, see appendix~\ref{apppropkasteleyn}.

\subsubsection{The positivity condition}
We now discuss the {\it positivity} condition.
The condition depends crucially on how the vertices of the graph are labeled.
It may appear to be merely a question of nomenclature, but it is more than
that: the labeling determines the order of the rows in the Pfaffian and
thereby affects its overall sign. We define the notion in precise terms:
a \defd{labeling} $\omega$ of the vertices of $g$ is a bijection
from $\mathcal V(g)$ to $\{1,\cdots,|g|\}$.

Given a vertex labeling $\omega$, a
pure dimer covering $\sigma\in \Omega_0(g)$, which we write as
$$\sigma=\{(v_1,v_2),\cdots,(v_{|g|-1},v_{|g|})\}$$
with $v_{2i-1}\succ v_{2i}$, is said to be \defd{positive} if the
permutation
$\pi^{(\omega)}_\sigma\in\mathcal S_{|g|}$ defined by
$\pi^{(\omega)}_\sigma(i)=\omega(v_i)$ has a positive signature. Note that
the sign of $\sigma$ remains unchanged if $(v_{2i-1},v_{2i})$
and $(v_{2j-1},v_{2j})$ are exchanged.
\begin{definition}\label{defpositivity}
Given a vertex labeling $\omega$, a directed graph $g$ is said to be
\defd{positive} if all of its dimer coverings are positive, or if it has no dimer coverings.
\end{definition}

The following proposition is the basis of Kasteleyn's theorem \cite{Ka63}.

\begin{proposition}[Uniform positivity]\label{propkasteleyn}
 Given a vertex labeling $\omega$, a Kasteleyn graph $g$ that admits a dimer covering is positive if and
only if one of its dimer coverings is positive.
\end{proposition}

Note that, in light of this proposition, every non-positive labeling can be made positive by 
switching two labels. 

\bigskip

We are finally in the position of stating Kasteleyn's theorem. Given a positive Kasteleyn graph $g\in\mathcal G$, let, for $i,j=1,\cdots,|g|$ with $i<j$,
\begin{equation}
 a_{i,j}(\mathbf d):=\left\{
 \begin{array}{l@{\ }l}
  +d_{v_i,v_j}&\mathrm{if\ }\{v_i,v_j\}\in\mathcal E(g)\ \mathrm{and\ }v_i\succ v_j\\
  -d_{v_i,v_j}&\mathrm{if\ }\{v_i,v_j\}\in\mathcal E(g)\ \mathrm{and\ }v_i\prec v_j\\
  0&\mathrm{otherwise}
 \end{array}
 \right.
\label{eqadef}\end{equation}
in which $v_i$ is a shorthand for
$\omega^{-1}(i)$. Proposition~\ref{propkasteleyn} implies that the terms
in
the Pfaffian $\mathrm{pf}(a(\mathbf d))$ (see~(\ref{eqpfaffdef})) add up
constructively, which in turn implies the following

\begin{theorem}[Kasteleyn's theorem]\label{theokasteleyn}
Given a positive Kasteleyn graph $g\in\mathcal G$, the partition function $\Xi(0,\mathbf d)$ of pure dimer coverings of $g$ is given by
 \begin{equation}
  \Xi(0,\mathbf d)=\mathrm{pf}(a(\mathbf d)).
 \label{eqkasteleyntheo}\end{equation}
\end{theorem}
\medskip

Note that, as commented above, every planar graph can be directed and labeled so as to make it Kasteleyn and positive. 
We remark that there are several directing procedures and labelings that ensure the Kasteleyn and positivity conditions. 
In the following, we are interested in proving that the sub-graphs obtained by erasing some vertices on the boundary 
(those at the monomer locations) are also Kasteleyn and positive. The subtle property to prove is the positivity of all such sub-graphs,
which is false in general. 
The goal of this paper is to find one good labeling of the full graph, guaranteeing positivity of all these sub-graphs. 

\subsection{A theorem on Pfaffians}\label{subsecpfaffian}

The basic strategy to prove our main result is to combine Kasteleyn's theorem with a general theorem
about Pfaffians \cite{Li68}, proved by one of us, which appears
at first glance to compute the
full MD partition function but fails to do so because of sign problems. Our
goal will be to show that these sign problems can be dealt with, if one
restricts the monomer locations to be on the boundary of a planar
graph, by making a careful choice of the direction and labeling of $g$.

\subsubsection{Statement of the theorem on Pfaffians}
We first state the theorem on Pfaffians, which is a slight generalization
of that proved in \cite{Li68}.

\begin{theorem}[Lieb]\label{theolieb}
Given an even positive integer $N$, an antisymmetric $N\times N$ matrix $a$,
and a collection of weights $\bm\ell=(\ell_i)_{i=1,\cdots,N}$, let
\begin{equation}
A_{i,j}(\bm \ell):=a_{i,j}-(-1)^{i+j}\ell_i\ell_j
\label{eqpolystal}\end{equation}
for $i<j$ and $A_{i,j}(\bm \ell)=-A_{j,i}(\bm \ell)$ for $i>j$, we have
\begin{equation}
\mathrm{pf}(A(\bm \ell))=\sum_{k=0}^{N/2}\sum_{\displaystyle\mathop{
\scriptstyle\mathcal
I\subset\{1,\cdots,N\}}_{|\mathcal
I|=2k}}\mathrm{pf}(\left[a\right]_{\mathcal I})\prod_{i\in\mathcal
I}\ell_i
\label{eqliebtheo}\end{equation}
in which
$\left[a\right]_{\mathcal I}$ denotes the matrix obtained from $a$ by
removing the $i$-th line and column for every $i$ in $\mathcal I$, and if $\mathcal I=\{1,\cdots,N\}$, then $\mathrm{pf}([a]_{\{1,\cdots,N\}})\equiv1$.
\end{theorem}
\medskip

In \cite{Li68}, the theorem was proved in the case in which the $\ell_i$
are equal, but the proof is immediately generalizable to arbitrary
$\bm \ell$. The only change needed in the proof of \cite{Li68} is to change
equation \cite[(21)]{Li68} from
\begin{equation}
\frac1Z\mathrm{trace}\left(\prod_{i=1}^N(\lambda+C_i)\right)
\qquad\mathrm{to}\qquad
\frac1Z\mathrm{trace}\left(\prod_{i=1}^N(\ell_i+C_i)\right).
\label{eqreplLi}\end{equation}
The rest of the proof is identical.
\bigskip

If we let all the $\ell_i$'s
equal $z$, then $\mathrm{pf}(A(\bm \ell))$ in~(\ref{eqliebtheo}) is the
polynomial in $z$ whose $2k$-coefficient is the sum of all sub-Pfaffians of
order $2k$. If $a_{i,j}$ is defined as in~(\ref{eqadef}), this {\it seems}
to count all MD coverings with $2k$ monomers: indeed, by Kasteleyn's
theorem, $\mathrm{pf}([a]_{\mathcal I})$ {\it appears} to be the partition
function of dimer coverings of the graph $[g]_{\mathcal I}$ obtained from
$g$ by removing the vertices whose labels are in $\mathcal I$, or
equivalently of MD coverings with monomers on the vertices whose labels are
in $\mathcal I$. This is not the case, however, {\it since $[g]_{\mathcal I}$ is
not necessarily a positive Kasteleyn graph}.

In the rest of this paper, we will provide an algorithm to direct and label
$g$ in such a way that when the vertices in $\mathcal I$ are restricted to
the boundary, which is imposed by setting all other $\ell_i$'s to zero,
$[g]_{\mathcal I}$ is a positive Kasteleyn graph. In that case,
$\mathrm{pf}(A(\bm\ell,\mathbf d))$ is the boundary MD partition function
with $A(\bm\ell,\mathbf d)$ defined in~(\ref{eqpolysta}).

\subsubsection{Lower bound on the monomer-dimer partition function}
When $\mathrm{pf}(A(\bm\ell,\mathbf d))$ does {\it not} equal the MD partition
function, it is so either because the terms in a sub-Pfaffian
$[a(\mathbf d)]_{\mathcal I}$ do not add up constructively, or because the sign of
$\mathrm{pf}([a(\mathbf d)]_{\mathcal I})$ is wrong. Nevertheless, the following
theorem holds.

\begin{theorem}[Lower bound for the terms in the MD partition
function]\label{theolowerbound}
For every $g\in\mathcal G$, if $d_e\ge0$ for all $e\in\mathcal
E(g)$, then defining
\begin{equation}
 a_{i,j}(\mathbf d):=\left\{
 \begin{array}{l@{\ }l}
  +d_{v_i,v_j}&\mathrm{if\ }\{v_i,v_j\}\in\mathcal E(g)\ \mathrm{and\ }v_i\succ v_j\\
  -d_{v_i,v_j}&\mathrm{if\ }\{v_i,v_j\}\in\mathcal E(g)\ \mathrm{and\ }v_i\prec v_j\\
  0&\mathrm{otherwise}
 \end{array}
 \right.
\label{eqAsimpledef}\end{equation}
in which $v_i$ is a shorthand for $\omega^{-1}(i)$ and, for $i<j$,
\begin{equation}
A_{i,j}(\bm\ell,\mathbf d):=a_{i,j}(\mathbf d)-(-1)^{i+j}\ell_i\ell_j,
\label{eqpolystasimple}\end{equation}
and $A_{j,i}(\bm\ell,\mathbf d):=-A_{i,j}(\bm\ell,\mathbf d)$, the Pfaffian $\mathrm{pf}(A(\bm\ell,\mathbf d))$ is a polynomial in the monomer weights
$\bm\ell$, each of whose coefficients are smaller or equal in absolute value  to the
corresponding term in the MD partition function $\Xi(\bm\ell,\mathbf d)$.
In other words, the coefficient of $\ell_{i_1}\cdots\ell_{i_k}$ is a lower
bound for the number of dimer coverings with monomers at $i_1,\cdots,i_k$.
\end{theorem}
\bigskip

\noindent {\bf Remark}: An upper bound for the MD partition function is provided in theorem~\ref{theoupperbound}.

\subsubsection{Proof of corollary \ref{corrmain}}
Corollary~\ref{corrmain} follows easily from theorems~\ref{theomain} and~\ref{theolieb}. Indeed, by~(\ref{eqtheo}) and~(\ref{eqliebtheo}),
\begin{equation}
M_n(i_1,\cdots,i_{2n})=
\frac{\mathrm{pf}([a(\mathbf d)]_{\mathcal I})}{\mathrm{pf}(a(\mathbf d))}
\label{eqpfM}\end{equation}
with $\mathcal I:=\{i_1,\cdots,i_{2n}\}$. We then make use of the following result: given an invertible $2N\times 2N$ antisymmetric matrix $X$ and a set $s\subset\{1,\cdots,2N\}$ of even cardinality, denoting the sub-matrix of $X^{-1}$ obtained by keeping only the lines and columns  indexed by elements of $s$ by $\{X^{-1}\}_{s}$, we have
\begin{equation}
\frac{\mathrm{pf}([X]_{s})}{\mathrm{pf}(X)}=(-1)^{|s|/2}\mathrm{pf}\{X^{-1}\}_{s},
\label{eqpfminors}\end{equation}
which can easily be proved by block-diagonalizing $X$ via a special unitary transformation, in such a way that each block is a $2\times2$ matrix of the form 
$\begin{pmatrix} 0 &\alpha_i\\ -\alpha_i&0\end{pmatrix}$. It follows from~(\ref{eqpfminors}) that
\begin{equation}\begin{array}{r@{\ }>{\displaystyle}l}
M_n(i_1,\cdots,i_{2n})=&
(-1)^n\mathrm{pf}(\{a^{-1}(\mathbf d)\}_{\mathcal I})\\[0.3cm]
=&\frac{1}{2^nn!}\sum_{\pi\in\mathcal S_{2n}}(-1)^\pi\prod_{j=1}^n(-a^{-1}(\mathbf d))_{i_{\pi(2j-1)},i_{\pi(2j)}},
\end{array}\label{eqpfMconc}\end{equation}
which concludes the proof, by noting that
\begin{equation}
(-a^{-1}(\mathbf d))_{i,j}=M_1(i,j).
\label{eqM1a}\end{equation}
\qed
\bigskip

\noindent{\bf Remark}: We have shown that theorem~\ref{theomain} implies corollary~\ref{corrmain}. It turns out that the converse is also true assuming~(\ref{eqM1a}) holds (with the sign that appears in~(\ref{eqM1a})). More precisely, corollary~\ref{corrmain} and~(\ref{eqM1a}) imply theorem~\ref{theomain}. Indeed, (\ref{eqM1a}) and corollary~\ref{corrmain} imply
$$
\Xi_\partial(\bm\ell,\mathbf d)=\Xi_\partial(\mathbf0,\mathbf d)\sum_{\mathcal I\subset\{1,\cdots,|g|\}}\mathrm{pf}(\{-a^{-1}\}_{\mathcal I})\prod_{j\in\mathcal I}\ell_j
$$
which, by~(\ref{eqkasteleyntheo}), (\ref{eqpfminors}) and theorem~\ref{theolieb}, yields~(\ref{eqtheo}). As a consequence, if one were to prove the Wick rule for boundary monomers (possibly by extending the analysis of~\cite{GBK78} to the monomer-dimer problem) then the Pfaffian formula~(\ref{eqtheo}) could be recovered by directing and labeling the graph $g$ in such a way that~(\ref{eqM1a}) holds, which can be achieved by ensuring that $[g]_{\{v,v'\}}$ is positive and Kasteleyn for every $v,v'\in\mathcal V(\partial g)$ (in the present paper, we prove the Pfaffian formula~(\ref{eqtheo}) without first proving the Wick rule, and ensuring that $[g]_{\mathcal I}$ is Kasteleyn and positive for every $\mathcal I\subset\mathcal V(\partial g)$, which does not seem to be harder than proving it for sets of cardinality 2). In other words, the Pfaffian formula~(\ref{eqtheo}) that counts MD coverings with any number of monomers on the boundary can be seen as a consequence of a similar Pfaffian formula for the MD coverings with 2 monomers on the boundary and the Wick rule.

\section{Proof of the main result for enclosed graphs}\label{secenclosed}

We first consider a class of graphs, called the set of {\it enclosed graphs}, that have a boundary circuit, and will subsequently show how to reduce any graph to such a case.
\bigskip

\begin{figure}
\hfil\includegraphics[width=6cm]{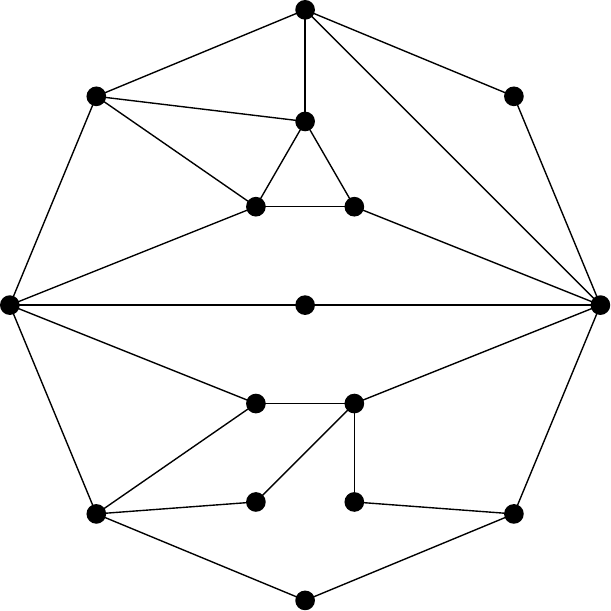}
\caption{An enclosed graph.}
\label{figenclosedexample}\end{figure}

The family of \defd{enclosed graphs} $\mathcal G_{en}\subset\mathcal G$ is
defined in the following way. Enclosed graphs are connected and consist of an interior graph $\check g\in\mathcal G$ (which may be empty) enclosed within a boundary circuit of vertices $\partial g$ containing an {\it even} number $|\partial g|$ of vertices. Since enclosed graphs are connected, $\partial g$ must be connected to each connected component of $\check g$ by at least one edge. In addition, vertices of $\partial g$ may be connected to each other by internal edges. See figure~\ref{figenclosedexample} for an example. Note that since $|g|$ is even, $|\check g|$ is even as well.
\bigskip

In the following, we will need to generalize the notion of positivity of a dimer covering to boundary monomer-dimer coverings (hereafter abbreviated as ``bMD coverings''). A bMD covering of a graph $g$ is said to be \defd{positive} if and only if the corresponding dimer covering of the graph obtained by removing the vertices occupied by monomers is positive.
\bigskip

\begin{figure}
\hfil\includegraphics[width=6cm]{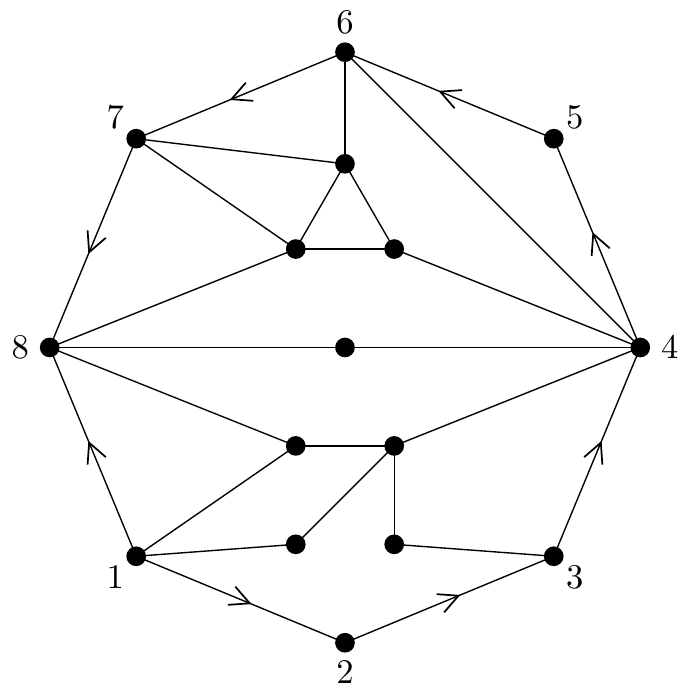}
\hfil\includegraphics[width=6cm]{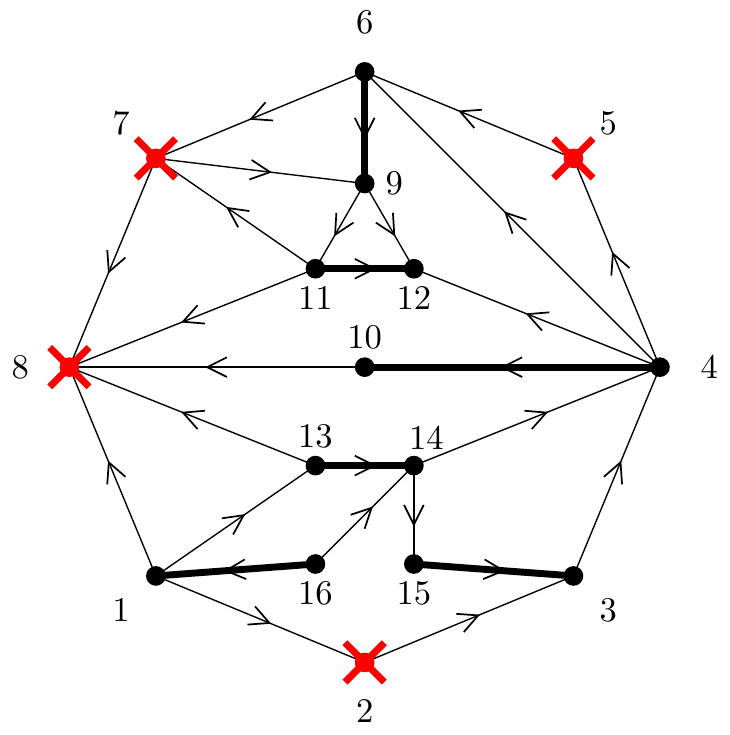}
\caption{Two stages of the directing and labeling of the graph in figure~\ref{figenclosedexample}. The first represents $g$ after having labeled the vertices of $\partial g$. The second shows a bMD of $g$  with monomers at 2, 5, 7, 8, and an associated labeling. The labels are chosen in such a way that this covering is positive:
$(16,1)$, $(15,3)$, $(4,10)$, $(6,9)$, $(11,12)$, $(13,14)$.}
\label{figenclosedexamplecover}\end{figure}

We will now describe how to direct and label an enclosed graph in such a way that its boundary MD partition function can be written as a Pfaffian as in theorem~\ref{theomain}. See figure~\ref{figenclosedexamplecover} for an example.
\bigskip

\delim{Directing and labeling an enclosed graph}
We first label the vertices of $\partial g$ following the edges of $\partial g$ sequentially in the counterclockwise direction. The resulting labeling is denoted by $\omega$. The location of the vertex labeled as 1 is unimportant.

We then direct the edges of $\partial g$:
$\omega^{-1}(i)\succ\omega^{-1}(i+1)$ and $\omega^{-1}(1)\succ\omega^{-1}(|\partial g|)$.
This implies that $\partial g$ is good. The
remaining edges of $g$ can be directed by
proposition~\ref{propdirectkasteleyn}. We arbitrarily choose one of the directions  constructed in its proof, see appendix~\ref{apppropkasteleyn}.

Finally, we label the remaining vertices in such a way that the resulting labeled graph is positive, by considering a random labeling, checking its sign, and exchanging two labels if it is negative, thus ensuring its positivity. The sign of the labeling can be computed either by constructing a dimer covering (or, more in general, a boundary monomer-dimer covering) and computing its sign, or by setting all weights $\ell_i=d_i=1$ and computing the sign of the right side of the Pfaffian formula~(\ref{eqtheo}).
\enddelim
\bigskip

If the monomers of an MD covering are fixed on vertices of $\partial
g$, the possible dimer
positions are the possible (pure) dimer coverings of the sub-graph of $g$
obtained by removing the vertices that have a monomer. We will now prove
that such a sub-graph is Kasteleyn and positive which implies that the dimer
partition function on it satisfies the Pfaffian
formula~(\ref{eqkasteleyntheo}) which can be substituted
in~(\ref{eqliebtheo}) to obtain~(\ref{eqtheo}). This result is contained in the following lemma, from which theorem~\ref{theomain}, restricted to the case of enclosed graphs, follows.
\bigskip

Given a family
of monomers $\mathcal M\subset\mathcal V(\partial g)$ of even cardinality,
we
define $[g]_{\mathcal M}$ as the sub-graph of $g$ obtained by removing the
vertices in $\mathcal M$.

\begin{lemma}\label{lemmaenclosedpositivity}
For every $g\in\mathcal G_{en}$, directed and labeled as above, for all
$\mathcal M\subset\mathcal V(\partial g)$ of even cardinality $|\mathcal
M|$, the sub-graph
$[g]_{\mathcal M}$ whose vertices are labeled by $\omega$ is
Kasteleyn and positive.
\end{lemma}
\medskip

\underline{Proof}: We first notice that $[g]_{\mathcal M}$ is Kasteleyn
because, since the monomers are on $\partial g$, the minimal circuits of
$[g]_{\mathcal M}$ are minimal circuits of $g$. In order to prove its
positivity, we will first construct an auxiliary graph $\gamma$ which is Kasteleyn and positive, and exhibit a mapping $\lambda_\gamma$ from dimer coverings of $[g]_{\mathcal M}$ to dimer coverings of $\gamma$, which preserves the sign of the covering, from which we conclude that $[g]_{\mathcal M}$ is positive.
\medskip

\begin{figure}
\hfil\includegraphics[width=10cm]{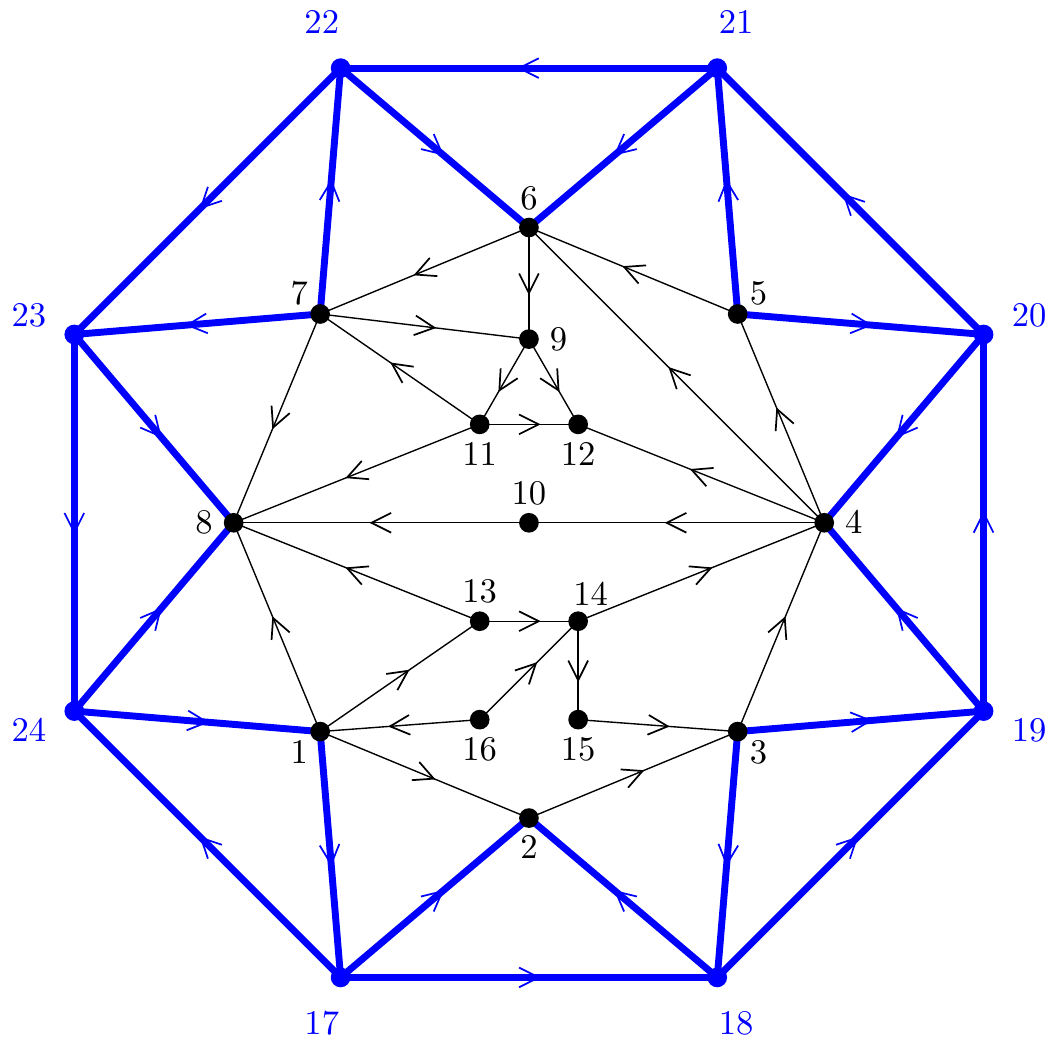}
\caption{The auxiliary graph associated to the graph in figure~\ref{figenclosedexample}. The extra edges are rendered as thick {\color{blue}blue} (color online) lines.}
\label{figenclosedauxiliary}\end{figure}

We construct $\gamma$ from $g$ in the following way (see figure~\ref{figenclosedauxiliary} for an example). We add an extra circuit of vertices $\epsilon$ outside of $g$, which consists of $|\partial g|$ vertices. The edges and vertices of $\gamma$ that are also edges or vertices of $g$ are directed and labeled in the same way as in $g$. We then label the vertices of $\epsilon$ sequentially in the counterclockwise direction from $|g|+1$ to $|g|+|\partial g|$ (the location of the first vertex is unimportant) and denote the resulting labeling by $\omega_\gamma$. We direct the edges of $\epsilon$ in such a way that $v\succ v'$ if and only if $\omega_\gamma(v)<\omega_\gamma(v')$, and add the following directed edges:
\begin{itemize}
\item $(\omega_\gamma^{-1}(2j-1),\omega_\gamma^{-1}(|g|+2j-1))$ and $(\omega_\gamma^{-1}(2j-1),\omega_\gamma^{-1}(|g|+2j-2))$ for $2\le j\le|\partial g|/2$
\item $(\omega_\gamma^{-1}(|g|+2j),\omega_\gamma^{-1}(2j))$ and $(\omega_\gamma^{-1}(|g|+2j-1),\omega_\gamma^{-1}(2j))$ for $1\le j\le|\partial g|/2$
\item $(\omega_\gamma^{-1}(1),\omega_\gamma^{-1}(|g|+1))$ and $(\omega_\gamma^{-1}(|g|+|\partial g|),\omega_\gamma^{-1}(1))$.
\end{itemize}
One readily checks that $\gamma$, thus directed, is Kasteleyn.
\medskip

\begin{figure}
\hfil\includegraphics[width=10cm]{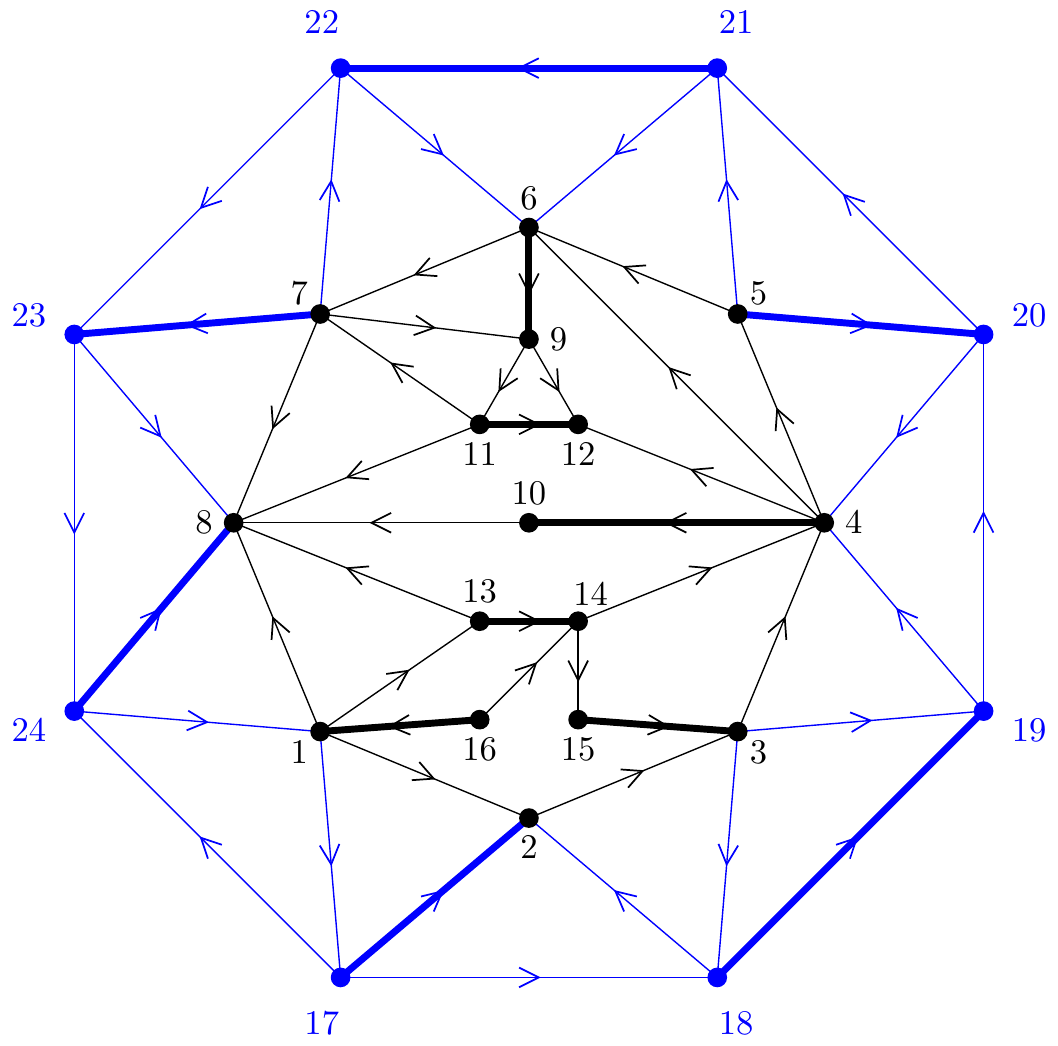}
\caption{The dimer covering associated to the bMD in figure~\ref{figenclosedexamplecover}.}
\label{figenclosedauxiliarycover}\end{figure}

We now construct an injective mapping $\lambda_{\gamma}$ from the set of bMD coverings of $g$ to the set of dimer coverings of $\gamma$. Given a bMD covering $\sigma$ of $g$, we construct $\lambda_\gamma(\sigma)$ in the following way (see figure~\ref{figenclosedauxiliarycover} for an example). We first add every dimer of $\sigma$ to $\lambda_\gamma(\sigma)$. For $1\le j\le|\partial g|$, let $p_j:=\mathrm{card}\{i< j\ |\ i\in\mathcal M\}$. For every $j\in\mathcal M$,
\begin{itemize}
\item if $j+p_j$ is odd, then we add $\{\omega_\gamma^{-1}(j),\omega_\gamma^{-1}(|g|+j)\}$ to $\lambda_\gamma(\sigma)$
\item if $j+p_j$ is even, then we add $\{\omega_\gamma^{-1}(j),\omega_\gamma^{-1}(|g|+j-1)\}$ to $\lambda_\gamma(\sigma)$.
\end{itemize}
In addition, for every $j\in\{1,\cdots,|\partial g|\}\setminus\mathcal M$,
\begin{itemize}
\item if $j+p_j$ is even, then we add $\{\omega_\gamma^{-1}(|g|+j),\omega_\gamma^{-1}(|g|+j-1)\}$ to $\lambda_\gamma(\sigma)$.
\end{itemize}
One readily checks that none of the dimers added in this way overlap, and that no vertices are left uncovered (in order to carry out the proof of this fact, it is convenient to consider the procedure described above from an algorithmic point of view, that is, considering each $j\in\{1,\cdots,|\partial g|\}$ successively, checking whether $j\in\mathcal M$ or not and the parity of $j+p_j$, and adding a dimer following the rules above; it is then straightforward to prove the property by induction on $j$).
\medskip

We will now prove that $\lambda_\gamma$ preserves the sign of $\sigma$. To do so, we first note that the two following edges are not occupied by a dimer of $\lambda_\gamma(\sigma)$:
\begin{itemize}
\item $(\omega_\gamma^{-1}(|g|+|\partial g|),\omega_\gamma^{-1}(1))$ (since if $1\in\mathcal M$, then $\{\omega^{-1}_\gamma(1),\omega_\gamma^{-1}(|g|+1)\}$ is a dimer of $\lambda_\gamma(\sigma)$),
\item $(\omega_\gamma^{-1}(|g|+1),\omega_\gamma^{-1}(|g|+|\partial g|))$ (by construction).
\end{itemize}
 This means that every every edge $\{v,v'\}$ with $v\in\mathcal V(\epsilon)$ and $v'\in\mathcal V(\partial g)$ that is occupied by a dimer of $\lambda_\gamma(\sigma)$ is directed as $v\succ v'$ if and only if $\omega_\gamma(v')$ is even, and that every edge $\{v,v'\}$ with $v,v'\in\mathcal V(\epsilon)$ that is occupied by a dimer and is directed $v\succ v'$ satisfies $\omega_\gamma(v')=\omega_\gamma(v)+1$.

We write the dimer covering $\lambda_\gamma(\sigma)$ as a sequence of labels
$$(j_1,j_2,\cdots,j_{|g|+|\epsilon|-1},j_{|g|+|\epsilon|})$$
in which
\begin{itemize}
\item $\{\omega_\gamma(j_{2i-1}),\omega_\gamma(j_{2i})\}$ is occupied by a dimer of $\lambda_\gamma(\sigma)$ and is oriented $\omega_\gamma(j_{2i-1})\succ\omega_\gamma(j_{2i})$,
\item $j_1,\cdots,j_{|g|-|\mathcal M|}\in\mathcal V(g)\setminus\mathcal M$
\item $\min\{j_{2i-1},j_{2i}\}<\min\{j_{2i'-1},j_{2i'}\}$ if and only if $i<i'$.
\end{itemize}
The sign of $\lambda_\gamma(\sigma)$ is the signature of the permutation that orders the indices $j_i$. We first focus on
$$(j_1,\cdots,j_{|g|-|\mathcal M|})$$
that is, on the indices corresponding to dimers of $g$. This sequence can be ordered by a permutation of signature $s$ where $s$ is the sign of $\sigma$. We denote the resulting ordered sequence by
$$(j'_1,\cdots,j'_{|g|-|\mathcal M|}).$$
We now focus on
$$(j_{|g|-|\mathcal M|+1},\cdots,j_{|g|+|\mathcal M|})$$
that is, on the indices corresponding to dimers covering edges between $\epsilon$ and $\partial g$. We first reorder these $j$'s in such a way that $j_{2i-1}\in\{1,\cdots,|\partial g|\}$ and $j_{2i}>|g|$ by a permutation whose signature is $(-1)^{\mathrm{card}\{i\in\mathcal M\ |\ i\ \mathrm{even}\}}$ (because of the way the edges were directed above). After this, it follows from the construction of $\lambda_\gamma(\sigma)$ that $j_{2i}<j_{2i'}$ for all $i<i'$, and from the definition of $j_i$, $j_{2i-1}<j_{2i'-1}$ for all $i<i'$. By a positive-signature permutation, we can therefore reorder the sequence to
$$(j'_{|g|-|\mathcal M|+1},\cdots,j'_{|g|+|\mathcal M|})$$
where $|\partial g|\ge j'_{|g|-|\mathcal M|+1}>\cdots>j'_{|g|}$ and $|g|<j'_{|g|+1}<\cdots<j'_{|g|+|\mathcal M|}$. Furthermore,
$$(j'_{|g|+1},\cdots,j'_{|g|+|\mathcal M|},j_{|g|+|\mathcal M|+1},\cdots,j_{|g|+|\epsilon|})$$
can be ordered by a positive signature permutation (since $j_{2i}=j_{2i-1}+1$ for $i>(|g|+|\mathcal M|)/2$). Finally,
$$(j'_1,\cdots,j'_{|g|-|\mathcal M|},j'_{|g|-|\mathcal M|+1},\cdots,j'_{|g|})$$
can be ordered by a permutation whose signature is $(-1)^{\mathrm{card}\{i\in\mathcal M\ |\ i\ \mathrm{odd}\}}$. By composing all of these permutations, we find that the sign of $\lambda_\gamma(\sigma)$ is
$$s(-1)^{\mathrm{card}\{i\in\mathcal M\ |\ i\ \mathrm{odd}\}}(-1)^{\mathrm{card}\{i\in\mathcal M\ |\ i\ \mathrm{even}\}}=s(-1)^{|\mathcal M|}=s.$$
Therefore, the sign of $\sigma$ and the sign of $\lambda_\gamma(\sigma)$ are equal.
\medskip

We can now conclude the proof of the lemma. By construction, one of the bMD coverings of $g$, which we denote by $\Sigma$, is positive, which implies that $\lambda_\gamma(\Sigma)$ is positive, and therefore, by proposition~\ref{propkasteleyn}, that $\gamma$ is positive. For every dimer covering $\sigma$ of $[g]_{\mathcal M}$, $\lambda_\gamma(\sigma)$ is, therefore, positive, which implies that $\sigma$ is positive, and concludes the proof of the lemma.\qed

\section{Reducing generic planar graphs to enclosed graphs}\label{sec7}
In this section, we discuss how a generic planar graph can be reduced to an enclosed graph. In particular, we will show how to add 0-weight edges to a graph to construct a boundary circuit, and how to add 0-weight edges and vertices to a graph that has an odd number of vertices in its boundary circuit to turn it into an enclosed graph, for which theorem~\ref{theomain} was proved in section~\ref{secenclosed}.

\subsection{Boundary circuit}\label{secboundarycircuit}
In this section, we give an algorithm to construct a boundary circuit for any planar graph $g$ by adding 0-weight edges. The construction is such that all the vertices of the boundary $\partial g$ of $g$ are in the boundary circuit. The boundary MD partition function on the graph with the boundary circuit is, therefore, equal to that on $g$, and the Pfaffian associated to the graph with the boundary circuit is equal to that associated to $g$.
\bigskip

First of all, if $g$ is disconnected, then we add 0-weight edges to it to connect its connected components to each other.

\begin{figure}
\hfil\includegraphics[width=6cm]{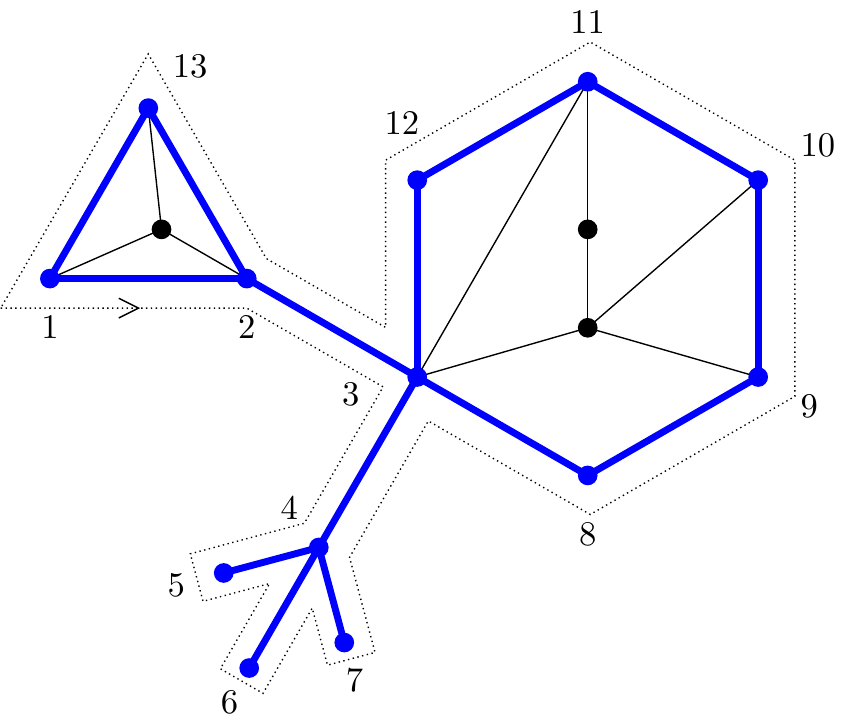}
\hfil\includegraphics[width=6cm]{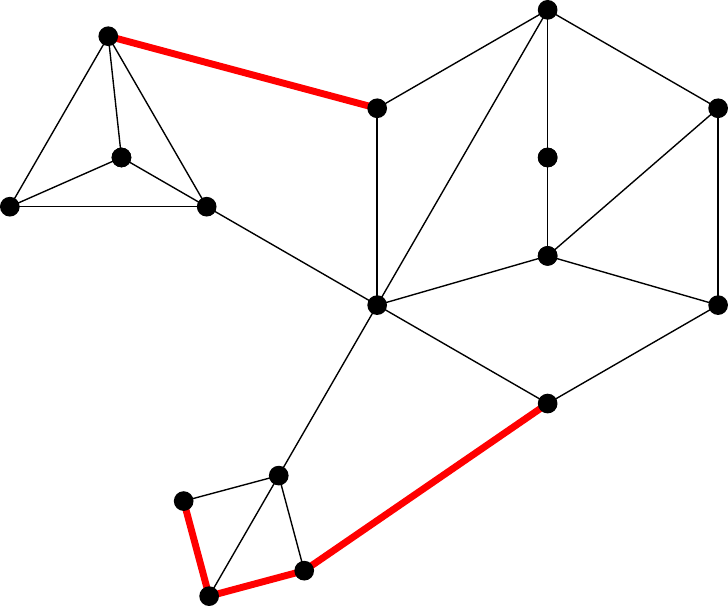}
\caption{A graph with no boundary circuit. In the leftmost figure, the edges and vertices of the boundary graph are drawn thicker and {\color{blue}blue}, and the dotted line represents the path shadowing the boundary from the outside, which we think of as starting and ending at $1$.  The corresponding string with repetitions is $(1,2,3,4,5,4,6,4,7,4,3,8,9,10,11,12,3,2,13)$.  After having erased the repetitions, we obtain the new string $(1,2,3,4,5,6,7,8,9,10,11,12,13)$. All the adjacent pairs but $\{5,6\}$, $\{6,7\}$, $\{7,8\}$ and $\{12,13\}$ are edges of $g$. In the rightmost figure, these four extra edges are added to the graph and drawn thicker and {\color{red}red}.}
\label{figboundaryexample}
\end{figure}

Then, consider a closed path shadowing the boundary of $g$ from the outside, and denote by $(v_1,\cdots,v_n)$ the string (possibly with repetitions)
listing the vertices of the boundary in the order encountered along the path. We identify $v_{n+1}\equiv v_1$. Then consider the ordered sub-set $(v_{i_1},\cdots,v_{i_k})$ of $(v_1,\cdots,v_n)$ obtained by 
erasing the repetitions. If a pair $\{v_{i_j},v_{i_{j+1}}\}$, $j=1,\ldots,k$, is not an edge of $g$, then we add a $0$-weight edge from $v_{i_j}$ to $v_{i_{j+1}}$. 
By construction, $(v_{i_1},\cdots,v_{i_k})$ is the boundary circuit of the resulting graph. See figure \ref{figboundaryexample} for an example. 

\bigskip

Once the boundary circuit has been constructed, if the boundary circuit contains an even number of vertices, then the resulting graph is an enclosed graph and can be directed and labeled as in section~\ref{secenclosed}. If the boundary circuit contains an odd number of vertices then the graph can be further reduced to an enclosed graph as explained in the following section.

\subsection{Making a boundary circuit even}\label{secevenboundarycircuit}
\begin{figure}
\hfil\parbox[m]{4cm}{\includegraphics[width=4cm]{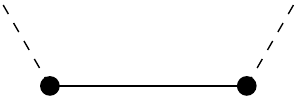}}
\hfil$\longmapsto$
\hfil\parbox[m]{4cm}{\includegraphics[width=4cm]{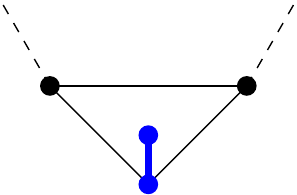}}
\caption{Replacing an edge of the boundary circuit in order to make it contain an even number of edges. The weight of the thick {\color{blue}blue} (color online) edge is set to 1. Note that since the {\color{blue} blue} edge must be occupied in every boundary MD covering, the weights of the other two extra edges and the weight of the extra vertex on the boundary circuit do not affect the boundary MD partition function.}
\label{figreplodd}\end{figure}

In this section, we show how to add edges and vertices to a graph with a boundary circuit that contains an odd number of vertices to an enclosed graph, in such a way that the boundary MD partition function is unchanged. The procedure is simple: pick an edge of the boundary circuit, and 
replace the edge according to figure~\ref{figreplodd}.

\bigskip

It is easy to check that, if the extra edge is given weight 1,  then the resulting graph has the same boundary MD partition function. In addition, the resulting graph is an enclosed graph, and can be directed and labeled as in section~\ref{secenclosed}, and it is straightforward to check that the Pfaffian computed from the graph with the extra edges and vertices is equal to that without. This concludes the proof of theorem~\ref{theomain}.
\vfill
\eject

\hfil{\Large\bf Acknowledgments}\par\penalty10000
\medskip
Thanks go to Jan Philip Solovej and Lukas Schimmer for devoting their time to some preliminary calculations that helped put us on the right track. We also would like to thank Tom Spencer and Joel Lebowitz for their hospitality at the IAS in Princeton and their continued interest in this problem. In addition, we thank Michael Aizenman and Hugo Duminil-Copin for discussing their work in progress on the random current representation for planar lattice models with us. We thank Jacques Perk and Fa Yueh Wu for very useful historical comments.

We gratefully acknowledge financial support from the A*MIDEX project ANR-11-IDEX-0001-02 (A.G.), from the PRIN National Grant {\it Geometric and analytic theory of Hamiltonian systems in finite and infinite dimensions} (A.G. and I.J.), and NSF grant PHY-1265118 (E.H.L.). 

\pagebreak

\appendix

\section{Properties of Kasteleyn graphs}\label{apppropkasteleyn}
In this appendix, we prove a few useful lemmas about Kasteleyn graphs. The key result is the following. 

\begin{lemma}\label{lemmacombcircuits}
Given two circuits $c_1$ and $c_2$ as in~(\ref{eqmergecircuit}), if $c_1$ and $c_2$ are good then their merger $c=c_1\Delta c_2$ is as well.
\end{lemma}
\medskip

\underline{Proof}: We write $c_1$ and $c_2$ as in~(\ref{eqmergecircuit}):
$$
c_1=(v_1,\cdots,v_{\nu+1},v_{\nu+2},\cdots,v_{|c_1|}),\quad
c_2=(v_{\nu+1},\cdots,v_{1},v_{\nu+2}',\cdots,v_{|c_2|}')
$$
and
$$
c=(v_{\nu+1},\cdots,v_{|c_1|},v_1,v_{\nu+2}',\cdots,v_{|c_2|}').
$$

We first prove that
\begin{itemize}
\item if $\nu$ is odd, then
  \begin{itemize}
  \item if $c_1$ and $c_2$ are either both oddly-directed or both evenly-directed then $c$ is
oddly-directed,
  \item otherwise $c$ is evenly-directed.
  \end{itemize}
\item if $\nu$ is even, then
  \begin{itemize}
  \item if $c_1$ and $c_2$ are either both oddly-directed or both evenly-directed then $c$ is evenly-directed,
  \item otherwise $c$ is oddly-directed.
  \end{itemize}
\end{itemize}

Indeed, a circuit $c_1$ is oddly-directed if and only if the numbers of backwards edges in
$(v_1,\cdots,v_{\nu+1})$ and in $(v_{\nu+1},\cdots,v_{|c_1|},v_1)$ have
different parity (by which we mean evenness or oddness), and
$c_2$ is oddly-directed if and only if the numbers of backwards edges in
$(v_{\nu+1},\cdots,v_{1})$ and in
$(v_1,v_{\nu+2}',\cdots,v_{|c_2|}',v_{\nu+1})$ have different parity.
Therefore, if $\nu$ is even, using the fact that the numbers of backwards
edges in $(v_1,\cdots,v_{\nu+1})$ and in $(v_{\nu+1},\cdots,v_1)$ have the
same parity, it follows that $c_1\Delta c_2$ is oddly-directed if and only if $c_1$ is
oddly-directed and $c_2$ is evenly-directed or vice-versa. If $\nu$ is odd, the numbers of
backwards edges in $(v_1,\cdots,v_{\nu+1})$ and in $(v_{\nu+1},\cdots,v_1)$
have different parity, so that $c_1\Delta c_2$ is oddly-directed if and only if $c_1$
and $c_2$ are either both oddly-directed or both evenly-directed.
\medskip

The proof of the lemma is then concluded by noticing that
if $\nu$ is odd then the number of vertices that are encircled by either
$c_1$ or $c_2$ has the same parity as the number of vertices encircled by
$c$, whereas the parity is different if $\nu$ is even.\qed
\bigskip

Lemma \ref{lemmacombcircuits} has a number of useful consequences, which are discussed in the following. 
\medskip

\noindent {\bf Proof of lemma~\ref{lemmakasteleyncomplete}}: Given a circuit $c$ of a Kasteleyn graph $g$, we prove
that it is good by induction on the number of edges it encloses. If it
is a minimal circuit (in particular if it encloses no edge), then it is
good by assumption. If not, then there exist $c_1$ and $c_2$ such that
$c=c_1\Delta c_2$, from which we conclude by the inductive hypothesis and
lemma~\ref{lemmacombcircuits}.\qed

\begin{lemma}\label{lemmakasteleynrmedge}
Given a Kasteleyn graph $g$, the graph obtained by removing any edge of $g$
is Kasteleyn.
\end{lemma}
\medskip

\underline{Proof}: The lemma follows from lemma~\ref{lemmakasteleyncomplete} and the fact that minimal circuits of the graph obtained by removing the edge are circuits of $g$.\qed
\bigskip

We will now prove proposition~\ref{propdirectkasteleyn} and, in the process, provide an algorithm to direct a planar graph $g$ in such a way that the resulting directed graph is Kasteleyn.
\medskip

\noindent{\bf Proof of proposition~\ref{propdirectkasteleyn}}: First of all, we notice that we can safely assume that $g$ has a boundary circuit: if it did not, then we construct an auxiliary graph $\gamma$ by adding edges to $g$, as detailed in section~\ref{secboundarycircuit}. Once $\gamma$ has been directed, the extra edges can be removed, and the Kasteleyn nature of the resulting directed graph then follows from lemma~\ref{lemmakasteleynrmedge}.

Assuming $g$ has a boundary circuit, we prove the proposition by induction on the number of
internal edges of the graph.

We first direct the edges of $\partial g$ in such a way that it is good
(if those edges are already directed then $\partial g$ is good by
assumption).

We first consider the case in which $\partial g$ is not a minimal circuit,
in which case there exist $c_1$ and $c_2$ such that $c_1\Delta c_2=c$. We
split $g$ into the graph $g_1$ consisting of $c_1$ and its interior and
$g_2$ consisting of $c_2$ and its interior. We direct the common edges
between $g_1$
and $g_2$ (or equivalently between $c_1$ and $c_2$) in such a
way that $c_1$ is good (there may be many ways of doing so, any one will do). By the inductive hypothesis, this implies that
$g_1$ can be directed appropriately. By lemma~\ref{lemmacombcircuits}, $c_2$
is good, which implies that $g_2$ can be directed as well.

We now turn to the case in which $\partial g$ is a minimal circuit (which
includes the case in which it has no interior edges).

If $\partial g$ encloses no circuit (that is if among the edges $\partial g$
encloses, if any, none form a circuit), then none of the edges enclosed in
$\partial g$ belong to a minimal circuit of $g$ (since that circuit would
have to contain an edge of $\partial g$). Therefore the edges enclosed in
$\partial g$ can be directed in any way without affecting the Kasteleyn
nature of $g$.

If $\partial g$ encloses at least one circuit, let $c_1,\cdots,c_n$ be the
maximal circuits enclosed by $\partial g$. The edges that are outside all of
the $c_i$'s do not belong to any minimal circuit and can therefore be
directed in any way. Let $g_i$ be the sub-graph of $g$ consisting of $c_i$
and its
interior. The sub-graph $g_i$ can be directed by the inductive
hypothesis.\qed
\bigskip

\section{Examples}\label{app.A}
In this appendix we provide some examples of Pfaffian formulas to illustrate the discussion.

\subsection{A graph with no interior vertices}
\begin{figure}
\hfil\parbox[m]{3cm}{\includegraphics[width=3cm]{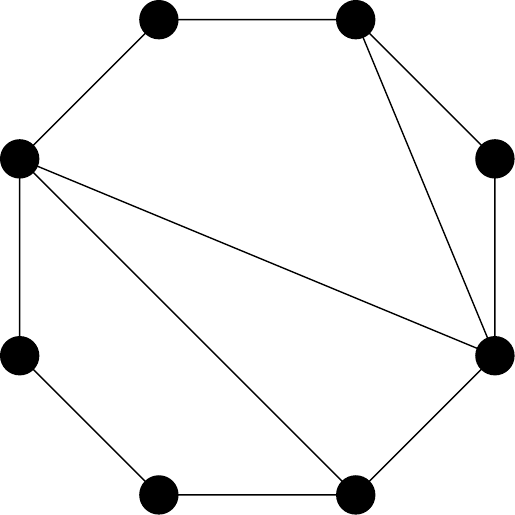}}
\hfil\parbox[m]{3.8cm}{\includegraphics[width=3.8cm]{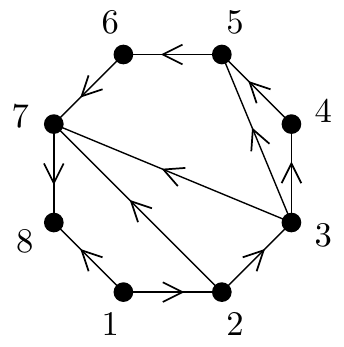}}
\caption{A graph with no interior vertices, and its direction and labeling.}
\label{figsimpleexample}\end{figure}

In this section, we compute the Pfaffian corresponding to figure~\ref{figsimpleexample}. We direct and label the graph as per the discussion in section~\ref{secenclosed} (see figure~\ref{figsimpleexample}). We set the edge weights $d_e=1$ and assume the monomer weights $\ell_i$ are all equal to $\sqrt{x}$. The antisymmetric matrix $A(\bm\ell,\mathbf d)$ is
$$
A(x)=\left(\begin{array}{cccccccc}
0&1+x&-x&x&-x&x&-x&1+x\\
&0&1+x&-x&x&-x&1+x&-x\\
&&0&1+x&1-x&x&1-x&x\\
&&&0&1+x&-x&x&-x\\
&&&&0&1+x&-x&x\\
&&&&&0&1+x&-x\\
&&&&&&0&1+x\\
&&&&&&&0
\end{array}\right)
$$
which is completed by antisymmetry. The MD partition function is therefore
\begin{equation}
\Xi(x)=\mathrm{pf}(A(x))=
x^4+11\,x^3+33\,x^2+28\,x+3.
\label{eqsimpleexample}\end{equation}

\subsection{Another graph with no interior: the L-shape}
\begin{figure}
\hfil\includegraphics{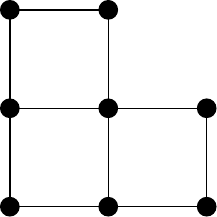}
\hfil\includegraphics{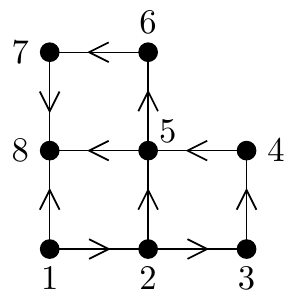}
\caption{The L-shape graph.}
\label{figLshape}
\end{figure}

In this section we compute the MD partition function for another simple graph: the {\it L-shape}, represented in figure~\ref{figLshape}. We direct and label the graph as per the discussion in section~\ref{secenclosed} (see figure~\ref{figLshape}) and find (setting $d_v=1$ and $\ell_i=\sqrt{x}$ as before)
$$
A(x)=\left(\begin{array}{cccccccc}
0&1+x&-x&x&-x&x&-x&1+x\\
&0&1+x&-x&1+x&-x&x&-x\\
&&0&1+x&-x&x&-x&x\\
&&&0&1+x&-x&x&-x\\
&&&&0&1+x&-x&1+x\\
&&&&&0&1+x&-x\\
&&&&&&0&1+x\\
&&&&&&&0
\end{array}\right)
$$
which is completed by antisymmetry. The MD partition function is therefore
\begin{equation}
\Xi(x)=\mathrm{pf}(A(x))=
x^4+10\,x^3+28\,x^2+24\,x+4
\label{eqLshape}\end{equation}

\subsection{A square graph}
\begin{figure}
\hfil\parbox[m]{3cm}{\includegraphics[width=3cm]{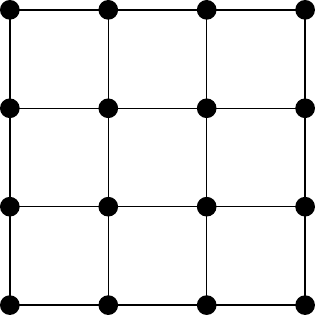}}
\hfil\parbox[m]{3.8cm}{\includegraphics[width=3.8cm]{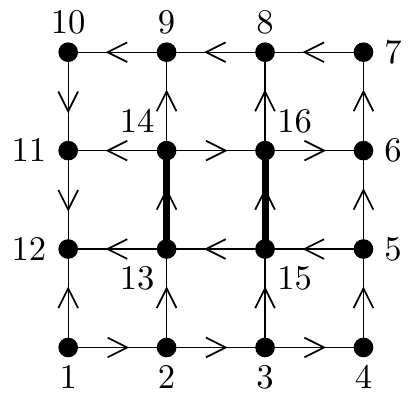}}
\caption{A square graph, and its direction and labeling.}
\label{figevenexample}\end{figure}

In this section, we compute the Pfaffian corresponding to the graph in figure~\ref{figevenexample}. We set $d_e=1$ and $\ell_i=\sqrt{x}$. Since the expression of the matrix $A$ is rather long, we split it into lines and only write the $i<j$ terms.
$$\begin{array}{r@{\ }r}
A_{ 1,\cdot}(x)=& 1+x,  -x,   x,  -x,   x,  -x,   x,  -x,   x,  -x, 1+x,   0,   0,   0,   0\\
A_{ 2,\cdot}(x)=& 1+x,  -x,   x,  -x,   x,  -x,   x,  -x,   x,  -x, 1  ,   0,   0,   0\\
A_{ 3,\cdot}(x)=& 1+x,  -x,   x,  -x,   x,  -x,   x,  -x,   x,   0,   0, 1  ,   0\\
A_{ 4,\cdot}(x)=& 1+x,  -x,   x,  -x,   x,  -x,   x,  -x,   0,   0,   0,   0\\
A_{ 5,\cdot}(x)=& 1+x,  -x,   x,  -x,   x,  -x,   x,   0,   0, 1  ,   0\\
A_{ 6,\cdot}(x)=& 1+x,  -x,   x,  -x,   x,  -x,   0,   0,   0,-1  \\
A_{ 7,\cdot}(x)=& 1+x,  -x,   x,  -x,   x,   0,   0,   0,   0\\
A_{ 8,\cdot}(x)=& 1+x,  -x,   x,  -x,   0,   0,   0,-1  \\
A_{ 9,\cdot}(x)=& 1+x,  -x,   x,   0,-1  ,   0,   0\\
A_{10,\cdot}(x)=& 1+x,  -x,   0,   0,   0,   0\\
A_{11,\cdot}(x)=& 1+x,   0,-1  ,   0,   0\\
A_{12,\cdot}(x)=&-1  ,   0,   0,   0\\
A_{13,\cdot}(x)=& 1  ,-1  ,   0\\
A_{14,\cdot}(x)=&   0, 1  \\
A_{15,\cdot}(x)=& 1  .
\end{array}$$
The MD partition function is therefore
\begin{equation}
\Xi(x)=\mathrm{pf}(A(x))=
2x^6 + 40x^5 + 256x^4 + 680x^3 + 776x^2 + 336x + 36.
\label{eqevenexample}\end{equation}

\subsection{An enclosed graph}
In this section, we compute the Pfaffian corresponding to the graph in figure~\ref{figenclosedexample}, directed and labeled as in figure~\ref{figenclosedexamplecover}. We set $d_e=1$ and $\ell_i=\sqrt{x}$. Since the expression of the matrix $A$ is rather long, we split it into lines and only write the $i<j$ terms.
$$\begin{array}{r@{\ }r}
A_{ 1,\cdot}(x)=& 1+x,  -x,   x,  -x,   x,  -x, 1+x,   0,   0,   0,   0, 1  ,   0,   0,-1  \\
A_{ 2,\cdot}(x)=& 1+x,  -x,   x,  -x,   x,  -x,   0,   0,   0,   0,   0,   0,   0,   0\\
A_{ 3,\cdot}(x)=& 1+x,  -x,   x,  -x,   x,   0,   0,   0,   0,   0,   0,-1  ,   0\\
A_{ 4,\cdot}(x)=& 1+x, 1-x,   x,-1-x,   0, 1  ,   0, 1  ,   0,-1  ,   0,   0\\
A_{ 5,\cdot}(x)=& 1+x,  -x,   x,   0,   0,   0,   0,   0,   0,   0,   0\\
A_{ 6,\cdot}(x)=& 1+x,  -x, 1  ,   0,   0,   0,   0,   0,   0,   0\\
A_{ 7,\cdot}(x)=& 1+x, 1  ,   0,-1  ,   0,   0,   0,   0,   0\\
A_{ 8,\cdot}(x)=&   0,-1  ,-1  ,   0,-1  ,   0,   0,   0\\
A_{ 9,\cdot}(x)=&   0, 1  , 1  ,   0,   0,   0,   0\\
A_{10,\cdot}(x)=&   0,   0,   0,   0,   0,   0\\
A_{11,\cdot}(x)=& 1  ,   0,   0,   0,   0\\
A_{12,\cdot}(x)=&   0,   0,   0,   0\\
A_{13,\cdot}(x)=& 1  ,   0,   0\\
A_{14,\cdot}(x)=& 1  ,-1  \\
A_{15,\cdot}(x)=&   0.
\end{array}$$
The MD partition function is therefore
\begin{equation}
\Xi(x)=\mathrm{pf}(A(x))=
22x^2 + 40x + 4.
\label{eqenclosedexample}\end{equation}

\subsection{A graph with no boundary circuit}
\begin{figure}
\hfil\includegraphics[width=6cm]{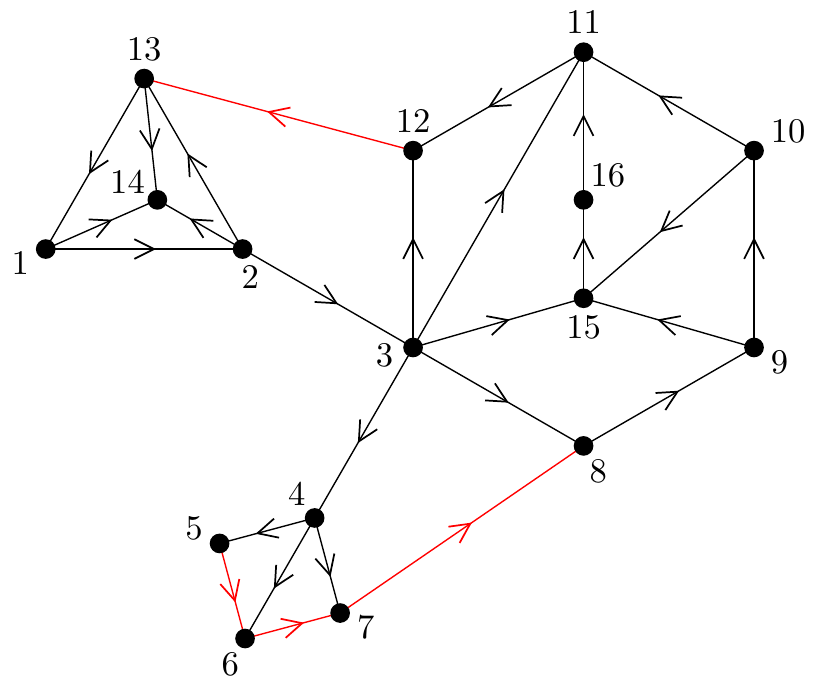}
\caption{Directing and labeling the graph in figure~\ref{figboundaryexample}.}
\label{figboundaryexampledir}
\end{figure}

In this section, we compute the Pfaffian corresponding to the graph in figure~\ref{figboundaryexample}, directed and labeled as in figure~\ref{figboundaryexampledir}. We set $d_e=1$ and $\ell_i=\sqrt{x}$. Since the expression of the matrix $A$ is rather long, we split it into lines and only write the $i<j$ terms.
$$\begin{array}{r@{\ }r}
A_{ 1,\cdot}(x)=& 1+x,  -x,   x,  -x,   x,  -x,   x,  -x,   x,  -x,   x,-1-x, 1  ,   0,   0\\
A_{ 2,\cdot}(x)=& 1+x,  -x,   x,  -x,   x,  -x,   x,  -x,   x,  -x, 1+x, 1  ,   0,   0\\
A_{ 3,\cdot}(x)=& 1+x,  -x,   x,  -x, 1+x,  -x,   x, 1-x, 1+x,  -x,   0, 1  ,   0\\
A_{ 4,\cdot}(x)=& 1+x, 1-x, 1+x,  -x,   x,  -x,   x,  -x,   x,   0,   0,   0\\
A_{ 5,\cdot}(x)=&   x,  -x,   x,  -x,   x,  -x,   x,  -x,   0,   0,   0\\
A_{ 6,\cdot}(x)=&   x,  -x,   x,  -x,   x,  -x,   x,   0,   0,   0\\
A_{ 7,\cdot}(x)=&   x,  -x,   x,  -x,   x,  -x,   0,   0,   0\\
A_{ 8,\cdot}(x)=& 1+x,  -x,   x,  -x,   x,   0,   0,   0\\
A_{ 9,\cdot}(x)=& 1+x,  -x,   x,  -x,   0, 1  ,   0\\
A_{10,\cdot}(x)=& 1+x,  -x,   x,   0, 1  ,   0\\
A_{11,\cdot}(x)=& 1+x,  -x,   0,   0,-1  \\
A_{12,\cdot}(x)=&   x,   0,   0,   0\\
A_{13,\cdot}(x)=& 1  ,   0,   0\\
A_{14,\cdot}(x)=&   0,   0\\
A_{15,\cdot}(x)=& 1 .
\end{array}$$
The MD partition function is therefore
\begin{equation}
\Xi(x)=\mathrm{pf}(A(x))=
3x^6 + 47x^5 + 222x^4 + 389x^3 + 234x^2 + 27x
\label{eqboundaryexample}\end{equation}

\section{An algorithm for the full monomer-dimer partition function}
\label{appalg}

In this appendix, we discuss an algorithm to compute the {\it full} MD partition function on an arbitrary graph (which is not necessarily planar).

The main idea is to isolate a {\it skeleton} $s$ from the graph, which is a sub-graph of $g$ obtained by removing edges from $g$ in such a way that $s$ is planar and contains no internal vertices. The boundary MD partition function of $s$ is the partition function of MD coverings of $g$ that does not have any dimers outside the skeleton. In order to count the coverings that do have dimers outside the skeleton, we add the following terms to the partition function. For every collection $\sigma$ of dimers that occupy edges that are outside the skeleton, we construct a sub-graph $[s]_\sigma$ of $s$ by removing the vertices covered by a dimer in $\sigma$. The boundary MD partition function of this sub-graph can be computed using theorem~\ref{theomain}. The full MD partition function is then obtained by summing the boundary MD partition functions of every such $[s]_\sigma$.

\bigskip

\begin{figure}
\hfil\includegraphics[width=5cm]{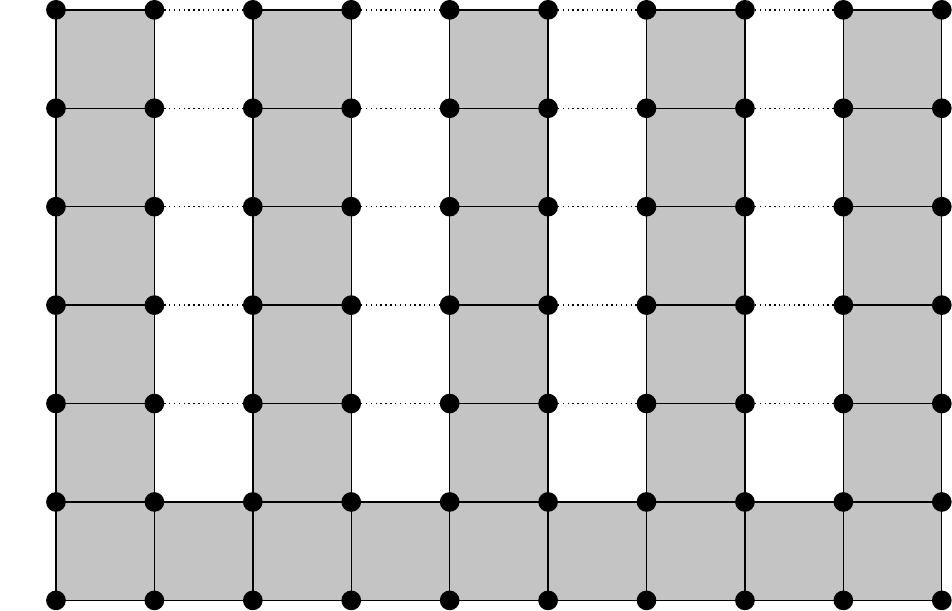}
\caption{A $10\times 7$ rectangle and its skeleton, colored grey.}
\label{figskeleton}
\end{figure}

If $g$ is an $L\times M$ sub-rectangle of $\mathbb Z^2$ with, say, $L$ even, then the skeleton can be constructed as in figure~(\ref{figskeleton}). By this algorithm, the MD partition function can be computed by summing $2^{\frac12(L-2)(M-2)}$ Pfaffians.

\begin{figure}
\hfil\includegraphics[width=3cm]{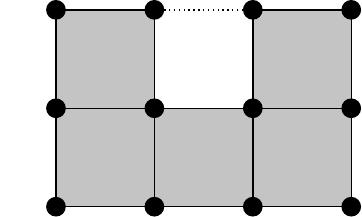}
\hfil\includegraphics[width=3cm]{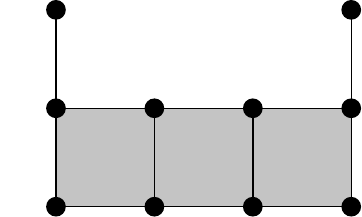}
\caption{The skeleton and its only sub-graph for the $3\times2$ rectangle.}
\label{figalgexample1}
\end{figure}

For example, if $L=4$, $M=3$ then, aside from the skeleton, there is a single sub-graph to be considered, see figure~\ref{figalgexample1}. The MD partition function is therefore the sum of two Pfaffians, which we have computed in the case $d_e=1$, $\ell_v=\sqrt{x}$:
\begin{equation}
\Xi(x)=
x^6 + 17x^5 + 102x^4 + 267x^3 + 302x^2 + 123x + 11.
\label{eqpfaff3x2}\end{equation}
If $L=6$, $M=6$, then the MD partition function is obtained by summing 256 Pfaffians:
\begin{equation}\begin{array}{r@{\ }l} 
\Xi(x)=&x^{18}+60 x^{17} + 1\,622\, x^{16} + 26\,172\, x^{15} + 281\,514 x^{14} +2\,135\,356\, x^{13}\\
&+ 11\,785\,382\, x^{12} + 
48\,145\,820\, x^{11} + 146\,702\,793\, x^{10} +
 333\,518\,324\, x^9\\
&+    562\,203\,148\, x^8 + 693\,650\,988\, x^7 + 613\,605\,045\, x^6 + 377\,446\,076\, x^5\\
&+ 154\,396\,898\, x^4 + 39\,277\,112\, x^3 +  5\,580\,152\, x^2 + 363\,536\, x +
 6\,728\;.
\end{array}\label{eqpfaff5x5}\end{equation}
Both~(\ref{eqpfaff3x2}) and~(\ref{eqpfaff5x5}) are in agreement with the results published in \cite[Table~6.7, column $N=12$]{Kr06} and \cite[Table~6.3]{Kr06} respectively.

\section{Another algorithm for the full monomer-dimer partition function}
\label{appinout}
If a graph $g\in\mathcal G$ is {\it Hamiltonian}, i.e. if there exists a circuit, called a {\it Hamiltonian cycle}, that goes through every vertex of $g$ exactly once, then we will now show how to write the {\it full} MD partition function on $g$ as a product of two Pfaffians. The condition that $g$ is Hamiltonian, is not restrictive, since 0-weight edges and vertices can be added to $g$ to make it so.
\medskip

Given a Hamiltonian cycle $c$, let $g_i$ denote the graph obtained from $g$ by removing every edge outside $c$ (that is the edges that are neither part of the Hamilton cycle, nor enclosed by it), and $\bar g_e$ the graph obtained from $g$ by removing every edge enclosed by $c$. We then consider a new embedding of $\bar g_e$, denoted by $g_e$, that is such that every vertex of $g_e$ is on the boundary (this is achieved by turning it inside out, that is, by setting the infinity-face of $g_e$ from the outside to the inside of the Hamilton cycle in $\bar g_e$).

The monomer dimer-partition function of $g$ can then be computed in the following way. We first set the weights of the edges and vertices of $g_e$ and $g_i$:
\begin{itemize}
\item given a vertex $v\in\mathcal V(g)$, we denote the weight of $v$ in $g_i$ by $\lambda_v$, and set the weight of $v$ in $g_e$ to the same value $\lambda_v$,
\item for every edge $e\in\mathcal E(c)$ that is part of the Hamilton cycle $c$, we denote the weight of $e$ in $g_i$ by $\delta_e$ and set the weight of $e$ in $g_e$ to the same value $\delta_e$,
\item for every edge $e\in\mathcal E(g)\setminus\mathcal E(c)$ that is {\it not} part of the Hamilton cycle $c$, $e$ is either an edge of $g_i$ or an edge of $g_e$; in either case, its weight is denoted by $\delta_e$.
\end{itemize}
Let $\Xi_i$ and $\Xi_e$ be the boundary MD partition functions on $g_i$ and $g_e$ respectively. The function $\Xi_i\Xi_e$ is a polynomial of order 2 in $\lambda_v$ and $\delta_e$. The terms in $\Xi_i\Xi_e$ that correspond to an MD covering of $g$ are those in which the corresponding coverings of $g_i$ and $g_e$ satisfy the following conditions:
\begin{itemize}
\item an edge $e\in\mathcal E(c)$ is occupied by a dimer in $g_i$ if and only if it is occupied in $g_e$ as well,
\item an edge $e=\{v,v'\}\in\mathcal E(g)\setminus\mathcal E(c)$ is occupied by a dimer in $g_i$ if and only if $v$ and $v'$ are occupied by monomers in $g_e$, and vice-versa,
\item a vertex $v\in\mathcal V(g)$ that is not covered by a dimer on $\mathcal E(g)\setminus\mathcal E(c)$, is occupied by a monomer in $g_i$ if and only if it is occupied in $g_e$ as well.
\end{itemize}
Therefore
\begin{equation}\begin{array}{r@{\ }>\displaystyle l}
\Xi(\bm\ell,\mathbf d)=&
\left(\prod_{v\in\mathcal V(g)}\left(1+\frac{\ell_v}2\frac{\partial^2}{\partial\lambda_v^2}\right)\right)
\left(\prod_{e\in\mathcal E(c)}\left(1+\frac{d_e}2\frac{\partial^2}{\partial\delta_e^2}\right)\right)\cdot\\[1cm]
&\left.\cdot\left(\prod_{e=\{v,v'\}\in\mathcal E(g)\setminus\mathcal E(c)}\left(1+d_e\frac{\partial^3}{\partial\delta_e\partial\lambda_v\partial\lambda_{v'}}\right)\right)
\Xi_i(\bm\lambda,\bm\delta)\Xi_e(\bm\lambda,\bm\delta)\right|_{\bm\lambda=0,\,\bm\delta=0}.
\end{array}\label{eqfullmdprodXi}\end{equation}
By theorem~\ref{theomain}, this implies the following theorem.

\begin{theorem}[Pfaffian formula for the full MD partition function]\label{theofullmd}
Given a Hamiltonian graph $g\in\mathcal G$, there exist two antisymmetric $|g|\times|g|$ matrices $A_i(\bm\lambda,\bm\delta)$ and $A_e(\bm\lambda,\bm\delta)$ such that
\begin{equation}\begin{array}{r@{\ }>\displaystyle l}
\Xi(\bm\ell,\mathbf d)=&
\left(\prod_{v\in\mathcal V(g)}\left(1+\frac{\ell_v}2\frac{\partial^2}{\partial\lambda_v^2}\right)\right)
\left(\prod_{e\in\mathcal E(c)}\left(1+\frac{d_e}2\frac{\partial^2}{\partial\delta_e^2}\right)\right)\cdot\\[1cm]
&\left.\cdot\left(\prod_{e=\{v,v'\}\in\mathcal E(g)\setminus\mathcal E(c)}\left(1+d_e\frac{\partial^3}{\partial\delta_e\partial\lambda_v\partial\lambda_{v'}}\right)\right)
\mathrm{pf}(A_i(\bm\lambda,\bm\delta))\mathrm{pf}(A_e(\bm\lambda,\bm\delta))\right|_{\bm\lambda=0,\,\bm\delta=0}.
\end{array}\label{eqfullmdprodpf}\end{equation}
\end{theorem}
\noindent The matrices $A_i$ and $A_e$ are constructed by directing and labeling $g_i$ and $g_e$ as in theorem~\ref{theomain}.
\bigskip

\noindent{\bf Remark}: It is important to note that this does not contradict the intractability result of M.~Jerrum~\cite{Je87}: indeed, (\ref{eqfullmdprodpf}) cannot, in general, be computed in polynomial-time. Indeed, since the entries of $A_i$ and $A_e$ are polynomials of $|g|+|\mathcal E(g)|$ variables, and computing their Pfaffian requires $O(|g|^3)$ multiplications of such elements, the computation of $\Xi$ via~(\ref{eqfullmdprodpf}) requires $O(|g|^32^{|g|+|\mathcal E(g)|})$ operations. This result extends to the Pfaffian formula in theorem~\ref{theomain}, but, there, if the weights $\ell_v$ and $d_e$ are given numerical values, or set to be equal among each other, the computation of the Pfaffian in~(\ref{eqtheo}) can be performed in polynomial-time. Because of the presence of derivatives in~(\ref{eqfullmdprodpf}), a similar operation cannot be done to compute~(\ref{eqfullmdprodpf}) in polynomial-time.
\bigskip

From theorem~\ref{theofullmd}, one can easily prove the following upper bound on the full MD partition function, which complements the lower bound in theorem~\ref{theolowerbound}:
\begin{theorem}[Upper bound for the terms in the MD partition
function]\label{theoupperbound}
Given a Hamiltonian graph $g\in\mathcal G$, there exist two antisymmetric $|g|\times|g|$ matrices $A_i(\bm\lambda,\bm\delta)$ and $A_e(\bm\lambda,\bm\delta)$ such that, if $d_e\ge0$ and $\ell_v>0$ for all $(v,e)\in\mathcal V(g)\times\mathcal E(g)$, the product
\begin{equation}
\mathrm{pf}(A_i(\bm\lambda,\bm\delta))\mathrm{pf}(A_e(\bm\lambda,\bm\delta))\Big|_{\begin{array}{>\scriptstyle l}
\lambda_v=\sqrt{\ell_v}\\
\delta_e=\sqrt{d_e}\,\mathrm{if}\,e\in\mathcal E(c)\\
\delta_e=\sqrt{d_e}\sqrt{\ell_v\ell_{v'}}^{-1}\,\mathrm{if}\,e=\{v,v'\}\not\in\mathcal E(c)
\end{array}}
\label{equpperbound}\end{equation}
is a Laurent polynomial in $\sqrt{\ell_v}$, each of whose coefficients are larger or equal to the corresponding term in the MD partition function $\Xi(\bm\ell,\mathbf d)$.
\end{theorem}

\section{The bijection method}\label{appbijectionmethod}

In this appendix, we show how to obtain an alternative Pfaffian formula for the boundary MD partition function via the {\it bijection method}. This construction was pointed out to us by an anonymous referee. It is related to the discussion in \cite[section 4]{Ku94}.
\bigskip

\subsection{Description of the method}
The main idea is to use the auxiliary graph $\gamma$ introduced in the proof of lemma~\ref{lemmaenclosedpositivity}, and show that the boundary MD partition function on $g$ is equal to half of the pure dimer partition function on $\gamma$, provided the edges of $\gamma$ are weighted appropriately. We set the weights of the edges of $\gamma$ in the following way:
\begin{itemize}
\item every edge of $\gamma$ that is also an edge of $g$ has the same weight as in $g$,
\item every edge of $\epsilon$ (see the proof of lemma~\ref{lemmaenclosedpositivity} for the definition of $\epsilon$) is assigned weight 1,
\item an edge between a vertex $v\in\mathcal V(\partial g)$ and a vertex $v'\in\mathcal V(\epsilon)$ is assigned the weight $\ell_v$.
\end{itemize}

We define a map $\Lambda_\gamma$ which maps a pure dimer covering of $\gamma$ to a bMD covering of $g$. Given a dimer covering $\Sigma$ of $\gamma$, we construct $\Lambda_\gamma(\Sigma)$ by putting monomers on the vertices of $\partial g$ that are occupied by a dimer of $\Sigma$ whose other end-vertex is in $\epsilon$, and by putting dimers on the edges of $g$ that are occupied by a dimer in $\Sigma$. Obviously, the weight of $\Sigma$ is equal to the weight of $\Lambda_\gamma(\Sigma)$.

Note that the map $\lambda_\gamma$ defined in the proof of lemma~\ref{lemmaenclosedpositivity} satisfies $\Lambda_\gamma(\lambda_\gamma(\sigma))=\sigma$ for every bMD covering $\sigma$ of $g$. Furthermore, we define another map $\bar\lambda_\gamma$ from the bMD coverings of $g$ to the dimer coverings of $\gamma$, similarly to $\lambda_\gamma$, but with $p_j$ replaced by $p_j+1$ (see the proof of lemma~\ref{lemmaenclosedpositivity}). This map also satisfies $\Lambda_\gamma(\bar\lambda_\gamma(\sigma))=\sigma$ for every bMD covering $\sigma$ of $g$. In addition, one easily checks that $\lambda_\gamma(\sigma)\neq\bar\lambda_\gamma(\sigma)$.
\bigskip

We wish to prove that for every bMD covering $\sigma$ of $g$, there are exactly two distinct pure dimer coverings $\Sigma_1$ and $\Sigma_2$ of $\gamma$ that satisfy $\Lambda_\gamma(\Sigma_i)=\sigma$. This is obvious if $\sigma$ has no monomers, so we will assume that $\sigma$ has at least one monomer, located on the vertex labeled as 1. Let $\sigma$ be such a covering. The coverings $\lambda_\gamma(\sigma)$ and $\bar\lambda_\gamma(\sigma)$ satisfy the required condition. One can then easily show, by induction, that having fixed a dimer on $\{\omega^{-1}_\gamma(1),\omega^{-1}_\gamma(|g|+1)\}$ as in $\lambda_\gamma(\sigma)$, $\lambda_\gamma(\sigma)$ is the {\it only} dimer covering of $\gamma$ that satisfies $\Lambda_\gamma(\lambda_\gamma(\sigma))=\sigma$. A similar argument can be made for $\bar\lambda_\gamma(\sigma)$. This implies that $\lambda_\gamma(\sigma)$ and $\bar\lambda_\gamma(\sigma)$ are the only dimer coverings of $\gamma$ satisfying $\Lambda(\Sigma_i)=\sigma$.

In conclusion, the bMD partition function on $g$ is equal to half of the pure dimer partition function on $\gamma$. By Kasteleyn's theorem, the bMD partition function can, therefore, be written as a Pfaffian.

\subsection{Example}

Let us look at a simple example and see how the Pfaffian formula one obtains from the bijection method differs from that presented in theorem~\ref{theomain}.

\begin{figure}
\hfil\parbox[m]{2cm}{\includegraphics[width=2cm]{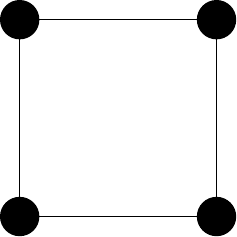}}
\hfil\parbox[m]{6cm}{\includegraphics[width=6cm]{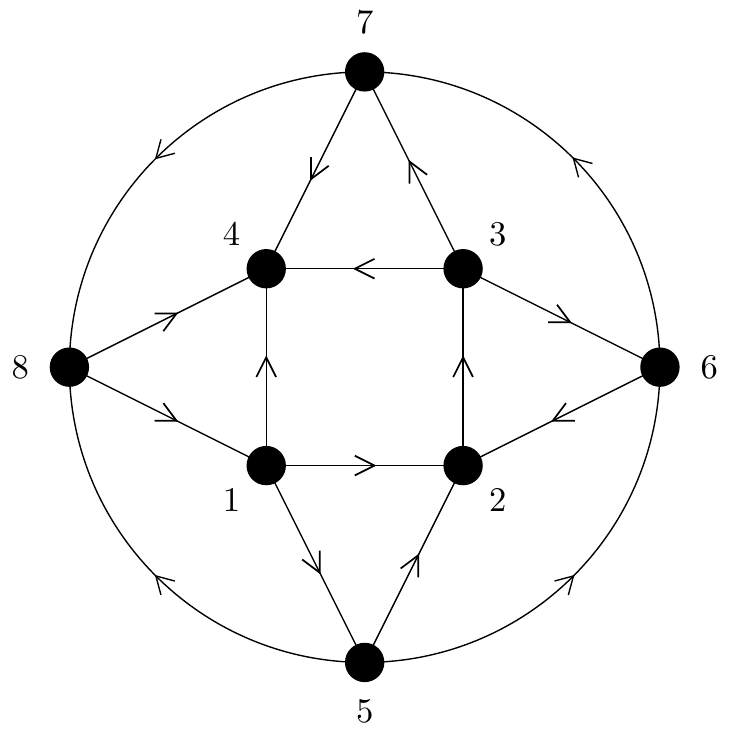}}
\caption{The square graph and the associated auxiliary graph $\gamma$.}
\label{figsquare}\end{figure}

Consider the square graph (see figure~\ref{figsquare}). Using the bijection method, we find that the bMD partition function on the square graph at dimer fugacity 1 and monomer fugacity $z$ is
\begin{equation}
\Xi_\partial=\frac12\mathrm{pf}\left(\begin{array}{cccccccc}
0 & 1& 0& 1& z& 0& 0&-z\\
-1& 0& 1& 0&-z&-z& 0& 0\\
 0&-1& 0& 1& 0& z& z& 0\\
-1& 0&-1& 0& 0& 0&-z&-z\\
-z& z& 0& 0& 0& 1& 0& 1\\
 0& z&-z& 0&-1& 0& 1& 0\\
 0& 0&-z& z& 0&-1& 0& 1\\
 z& 0& 0& z&-1& 0&-1& 0
\end{array}\right).\label{eqsquarebijection}\end{equation}
Using theorem~\ref{theomain}, we find
\begin{equation}
\Xi_\partial=\mathrm{pf}\left(\begin{array}{cccc}
   0& 1+z^2&  -z^2& 1+z^2\\
-1-z^2&   0& 1+z^2&  -z^2\\
   z^2&-1-z^2&   0& 1+z^2\\
-1-z^2&   z^2&-1-z^2&   0
\end{array}\right).\label{eqsquare}\end{equation}
Obviously, both formulas yield the same result:
\begin{equation}
\Xi_\partial=z^4+4z^2+2.
\label{eqsquareres}\end{equation}

\pagebreak

\end{document}